\documentclass[preprint,showpacs,preprintnumbers,amsmath,amssymb]{revtex4}
\usepackage{graphicx}
\usepackage{epsfig}
\usepackage{bm}
\usepackage{amsfonts}
\usepackage{subfigure}
\usepackage{multirow}
\usepackage{float}
\usepackage{color}

\providecommand{\tabularnewline}{\\}

\newcommand{\e}{\epsilon}

	\newcommand{\newc}{\newcommand}
	\newc{\be}{\begin{equation}}
		\newc{\ee}{\end{equation}}
	\newc{\ba}{\begin{eqnarray}}
		\newc{\ea}{\end{eqnarray}}
	\newc{\rH}{{\rm H}}
	\newc{\rd}{{\rm d}}
	\newc{\Mpl}{M_{\rm Pl}}

\begin{document}
	
	\title{\textbf{Primordial black holes in scalar field inflation coupled to the Gauss-Bonnet term with fractional power-law potentials}}

	
	\author{Ali Ashrafzadeh\footnote{a.ashrafzadeh@uok.ac.ir} and Kayoomars Karami\footnote{kkarami@uok.ac.ir}}
	\affiliation{\small{Department of Physics, University of Kurdistan, Pasdaran Street, P.O. Box 66177-15175, Sanandaj, Iran}}
	
	\date{\today}
\begin{abstract}

In this study, we investigate the formation of primordial black holes (PBHs) in a scalar field inflationary model coupled to the Gauss-Bonnet (GB) term with fractional power-law potentials. The coupling function enhances the curvature perturbations, then results in the generation of PBHs and detectable secondary gravitational waves (GWs).
We identify three separate sets of parameters for the potential functions of the form $\phi^{1/3}$, $\phi^{2/5}$, and $\phi^{2/3}$. By adjusting the model parameters, we decelerate the inflaton during the ultra slow-roll (USR) phase and enhance curvature perturbations.
Our calculations predict the formation of PBHs with masses of ${\cal O}(10)M_{\odot}$, which are compatible with LIGO-Virgo observational data. Additionally, we find PBHs with masses around ${\cal O}(10^{-6})M_{\odot}$ and ${\cal O}(10^{-5})M_{\odot}$, which can explain ultrashort-timescale microlensing events in OGLE data.
Furthermore, our proposed mechanism could lead to the formation of PBHs in mass scales around ${\cal O}(10^{-14})M_{\odot}$ and ${\cal O}(10^{-13})M_{\odot}$, contributing to approximately 99\% of the dark matter in the universe.
We also study the production of secondary GWs in our model. In all cases of the model, the density parameter of secondary GWs $\Omega_{\rm GW_0}$ exhibits peaks that intersect the sensitivity curves of GWs detectors, providing a means to verify our findings using data of these detectors.
Our numerical results demonstrate a power-law behavior for the spectra of $\Omega_{\rm GW_0}$ with respect to frequency, given by $\Omega_{\rm GW_0} (f) \sim (f/f_c)^{n}$. Additionally, in the infrared regime where $f\ll f_{c}$, the power index takes a log-dependent form, specifically $n=3-2/\ln(f_c/f)$.
\end{abstract}
	
	\pacs{ }
	
	\maketitle
	
	\newpage
	\section{Introduction}
The first idea of generating PBHs from primordial curvature perturbations was proposed by Zel'dovich and Novikov~\cite{zeldovich:1967}. Further studies have showed that low-mass PBHs have completely evaporated in the early universe due to quantum evaporation, while more massive PBHs could remain until now \cite{Hawking:1974,Hawking:1976,Page:1976,carr:1975,carr:1974}.
The detection of GWs from a binary black hole merger by the LIGO-Virgo collaboration has brought attention to the PBHs \cite{Abbott:2016,Abbott:2016-a,Abbott:2016-b,Abbott:2017-a,Abbott:2017-b,Abbott:2017-c,Abbott:2017,Abbott:2019}.
PBHs could be considered as a suitable candidate for explaining all or a portion of the dark matter in the universe \cite{Nojiri:2017,Olmo:2011,Faraoni:2010,Capozziello:2011,Nojiri:2011,Cruz-Dombriz:2012,Hawking:1971,Chapline:1975,Ivanov:1994,Polnarev:1985,Teimoori-b:2021,Solbi-b:2021,Rezazadeh:2021,Solbi-a:2021,Hajkarim:2019}.
Since PBHs originate from the gravitational collapse of primordial density perturbations, they are not subject to the Chandrasekhar mass limitation. Therefore, PBHs differ from astrophysical black holes and they can have a wide range of masses. In this regard, PBHs with  mass scale around the ${\cal O}(10)M_{\odot}$, could be the source of GWs in the LIGO-Virgo observations. Additionally, PBHs with mass around ${\cal O}(10^{-5})M_{\odot}$,  could explain the ultrashort-timescale microlensing events observed in OGLE data \cite{OGLE}.
Moreover, PBHs within the mass range of asteroid masses, approximately ${\cal O}(10^{-16}-10^{-11})M_{\odot}$, could constitute the entire dark matter content of the universe \cite{Gould:1992,Alcock:2001,Teimoori:2021,Heydari:2021,Heydari:2022,Heydari:2023,Dalcanton:1994,Laha:2020,Ali-Haimoud:2017,Sato-Polito:2019,Wang:2018,Jacobs:2015}.

The generation of PBHs in the radiation dominated (RD) era requires a large amplitude of primordial curvature perturbations during the inflationary epoch.
When the superhorizon scales associated with large amplitudes become subhorizon during the RD era, gravitational collapse of the overdense regions generates PBHs.
In order to generate PBHs, the scalar power spectrum should increase to the order of $\mathcal{O}(10^{-2})$ at small scales.
On the other hand, the cosmic microwave background (CMB) measurements have revealed that the amplitude of the scalar power spectrum at the pivot scale $k_{*}=0.05  \mathrm{Mpc}^{-1}$ should be equal to $2.1 \times 10^{-9}$ \cite{akrami:2020}.
Hence, the scalar power spectrum would need to be enhanced by seven orders of magnitude at small scales to be suitable for PBH formation \cite{Mishra:2020,Fu:2019,Kawai:2021,Dalianis:2019,mahbub:2020,Kawaguchia:2022,Villanueva-Domingo:2021,Gao:2021,Lin:2020,Kamenshchik:2019,Zhang:2022}.

The re-entry of overdense regions into the horizon during the RD era can lead to the generation of secondary GWs simultaneous with the formation of PBHs. Therefore, the detection of secondary GWs signals provides a novel approach for indirectly detecting PBHs \cite{Bhattacharya:2023}.

One of the most popular approaches for increasing the scalar power spectrum is to generate a short USR phase. During this phase, the inflaton decelerates effectively, and consequently the curvature perturbations could increase significantly \cite{Ragavendra:2023}.
The USR region, which is necessary for the production of PBHs during inflation, can be created through various methods. This aim could be reached  both within the standard model of inflation and modified gravity frameworks.
In standard single-field inflationary models, if the potential has an inflection point, the velocity of inflaton is greatly reduced in the vicinity of inflection point. Consequently, during the USR phase, the scalar power spectrum can be enhanced to generate suitable overdense regions for PBH formation \cite{Mishra:2020,Dalianis:2019,mahbub:2020}.
Also, in the context of modified gravity, the conditions for entering the USR phase can be established by choosing a suitable coupling function and the fine tuning of the model parameters ~\cite{Fu:2019,Kawai:2021,Kawaguchia:2022,Villanueva-Domingo:2021,Gao:2021,Lin:2020,Kamenshchik:2019,Zhang:2022}.
For instance, in the non-minimal derivative coupling framework \cite{Fu:2019} and the non-canonical models \cite{Gao:2021,Lin:2020,Kamenshchik:2019}, the authors have successfully explained the production of PBHs by employing suitable coupling functions.

Recently, the possibility of PBHs formation in modified gravity model with GB coupling has been studied \cite{Kawai:2021,Zhang:2022,Kawaguchia:2022}.
Kawai et al. \cite{Kawai:2021}, could demonstrate that the PBH formation from natural potential in GB framework could justify dark matter contribution of the universe. Additionally, the authors have studied the secondary GWs generation as a possibility of indirectly detecting the PBHs.\\
Also, Zhang \cite{Zhang:2022} has studied PBHs in the $\alpha$-attractor E-model inflation with the Gauss-Bonnet coupling. This coupling enables them to  enhance the curvature perturbations at small scales and produce a notable abundance of PBHs and detectable secondary GWs. The author has also estimated the primordial non-Gaussianity and discussed its impact on PBHs and secondary GWs.
In addition, Kawaguchia et al. \cite{Kawaguchia:2022} have computed the power spectra of primordial curvature perturbations generated in GB corrected Higgs inflation. In this model, the inflaton field has a nonminimal coupling to gravity as well as a GB coupling. This study has revealed the existence of parameter spaces wherein PBHs could have been taken into account for  all dark matter content of the universe.

In this study, we specifically focus on PBHs formation and the fractional energy density of secondary GWs in scalar field inflation coupled to the GB term with fractional power-law potentials. Note that the fractional power-law potentials in the standard model of inflation do not agree with the Planck observations.

This paper is organized as follows. Sect. \ref{sec2} provides a brief overview of the GB gravity. In Sect. \ref{sec3}, we explain the used technique  to increase the amplitude of the curvature power spectrum at small scales to order ${\cal O}(10^{-2})$ in the GB gravity.
In Sect. \ref{sec4}, we compute the various masses and fractional abundances of PBHs. In Sect. \ref{sec5}, we calculate the energy spectra of induced GWs. Finally, the abbreviated results are listed in Sect. \ref{sec6}.
\section{Gauss-Bonnet Inflationary Model}
\label{sec2}
We consider the Einstein-Gauss-Bonnet action as follows \cite{Jiang:2013,Koh:2014,Guo:2010,Odintsov:2020,Gao:2020,ShahraeiniNozari:2022,RashidiNozari:2020,AziziNozari:2022}
	\be
	{\cal S}=\int {\rm d}^4 x \sqrt{-g} \left[\frac{\Mpl^{2}}{2}R
	-\frac{1}{2}g^{\mu \nu}\nabla_{\mu}\phi
	\nabla_{\nu}\phi-V(\phi)-\frac{\xi(\phi)}{2}R_{\rm GB}^{2}\right],
	\label{action1}
	\ee
where $R$ is the Ricci scalar, $\phi$ is the scalar field, $R^2_{\rm GB} \equiv R_{\mu\nu\rho\sigma}R^{\mu\nu\rho\sigma} - 4R_{\mu\nu}R^{\mu\nu} + R^2$ is the invariant scalar GB term and $\xi(\phi)$ is the coupling function between the scalar field and GB term. Here $\Mpl=(8\pi G)^{-1/2}$ is the reduced Planck mass.

Here we consider a homogeneous and isotropic universe which is defined by a spatially flat Friedmann-Lema\^{\i}tre-Robertson-Walker (FLRW) metric with the line element given by
	\be
	{\rm d}s^2=-{\rm d}t^2+a^2(t) \delta_{ij}{\rm d}x^i {\rm d}x^j\,,
	\label{metric1}
	\ee
	where $a$ is the scale factor. The Friedmann equations and the equation of motion can be derived by taking variation of the action~(\ref{action1}) with respect to the metric $g^{\mu \nu}$ and the scalar field $\phi$, respectively as follows \cite{Jiang:2013,Koh:2014,Guo:2010,Odintsov:2020,Gao:2020}
	\ba
	\label{bge1}
	&& 3{\Mpl^{2}}H^2 = \frac12\dot{\phi}^2+V(\phi)+12\dot{\xi}H^3 \;, \\
	\label{bge2}
	&& -2{\Mpl^{2}}\dot{H} =\dot{\phi}^2 - 4\ddot{\xi}H^2 - 4\dot{\xi}H\left(2\dot{H} - H^2\right)\,, \\
	\label{bge3}
	&& \ddot{\phi}+3H\dot{\phi} = - V_{,\phi} - 12\xi_{,\phi}H^2\left(\dot{H}+H^2\right) \;,
	\ea
where, $H\equiv \dot{a}/a$ is the Hubble parameter, dot denotes the time derivative and the subscript $({,\phi})$ demonstrates derivative with respect to $\phi$.
In the GB gravity, the slow-roll parameters are defined as follows  \cite{Jiang:2013,Koh:2014,Guo:2010,Odintsov:2020,Gao:2020}
	\ba
	\label{SRH}
    \e_1\equiv\frac{-\dot{H}}{H^2}\;,~~\e_2\equiv\frac{\ddot{\phi}}{H \dot{\phi}}\;,~~ \delta_1\equiv4\dot{\xi}H \;,~~\delta_2\equiv\frac{\dot{\delta_1}}{H \delta_1}\,.
	\ea
	%
	 Under the slow-roll conditions ($|\e_{i}|\ll 1$ and $|\delta_{i}|\ll 1$), the background Eqs.~(\ref{bge1})-(\ref{bge3}) reduce to
	\ba
	\label{sre1}
	&& H^2 \simeq \frac{\Mpl^{2}}{3} V \;,\\
	\label{sre2}
	&& -2{\Mpl^{2}}\dot{H} \simeq \dot{\phi}^2 + 4\dot{\xi}H^3 \,,\\
	\label{sre3}
	&& 3H\dot{\phi} \simeq -V_{,\phi} - \frac{12}{\Mpl^{4}}\xi_{,\phi}H^4 \,.
	\ea
Also the slow-roll approximation of the scalar power spectrum ${\cal P}_s(k)$ at the sound horizon crossing moment is given by \cite{Kawaguchia:2022,Felice:2011}
	\be
	{\cal P}_{s}(k) = \frac{H^2}{8\pi^2 Q_s c_s^3}\biggr|_{c_s k=a H}\,,
	\label{powerspectraS}
	\ee
	where
	\be
	Q_s=16\frac{\Sigma}{\Theta^2}Q_t^2+12Q_t\,,\qquad
	c_s^2=\frac{1}{Q_s}\left[ \frac{16}{a}\frac{\rd}{\rd t}
	\left(\frac{a}{\Theta}Q_t^2\right)-4c_t^2 Q_t\right]\,,
	\label{cs}
	\ee
	with
	\ba
	\label{cT}
	\Sigma=\frac{1}{2}\dot{\phi}^2-3 \Mpl^2 H^2+24 H^3 \xi_{,\phi}\dot{\phi}\,,\qquad\quad
	\Theta=\Mpl^2 H-6 H^2 \xi_\phi\dot{\phi}\,,\qquad\nonumber\\
	Q_t=\frac{1}{4}\left(-4H\xi_{,\phi}\dot{\phi}+\Mpl^2\right)\,,\qquad
	c_t^2=\frac{1}{4Q_t}\left(\Mpl^2-4\xi_{,\phi\phi}\dot{\phi}^2-4\xi_{,\phi}\ddot{\phi}\right)\,.
	\ea
    The necessary conditions to prevent the occurrence of ghost and Laplacian instabilities associated with scalar and tensor perturbations are as follows
	\be
	Q_s>0\,,\qquad c_s^2>0\,,\qquad
	Q_t>0\,,\qquad c_t^2>0\,.
	\label{avoidinstability}
	\ee
	%
The amplitude of scalar perturbations ${\cal P}_s(k)$ at the pivot scale $k_{*}=0.05~\rm Mpc^{-1}$ is ${\cal P}_s(k_{*}) = 2.1\times{10}^{-9}$, based on Planck 2018 data \cite{akrami:2020}. Under the slow-roll approximation with the help of Eqs. (\ref{sre1})-(\ref{sre3}), the slow-roll parameters (\ref{SRH}), can be expressed in terms of the potential and GB coupling function as \cite{Jiang:2013,Koh:2014}
	\ba
	\e_1 \simeq \frac{Q}{2} \frac{V_{,\phi}}{V}\,,\qquad\quad
	\e_2 \simeq-Q\left(\frac{V_{,\phi\phi}}{V_{,\phi}}-\frac{V_{,\phi}}{V}+\frac{Q_{,\phi}}{Q}\right)\,,\nonumber\\
	\delta_1 \simeq -\frac{4Q}{3} \xi_{,\phi}{V} \,,\qquad
	\delta_2 \simeq -Q\left(\frac{\xi_{,\phi\phi}}{\xi_{,\phi}}+\frac{V_{,\phi}}{V}+\frac{Q_{,\phi}}{Q}\right)\,,
	\label{SRV}
	\ea
	with $Q\equiv V_{,\phi}/V+4\xi_{,\phi}V/(3{\Mpl^{4}})$.\\
By using the definition of the slow-roll parameters, Eq.~(\ref{SRV}), the scalar spectral index $n_s$ reads \cite{Jiang:2013}
	%
	\ba
	n_s-1 \equiv \frac{\rd \ln {\cal P}_s}{\rd \ln k}\biggr|_{c_s k=aH} \simeq -2\e_1-\frac{2\e_1\e_2-\delta_1\delta_2}{2\e_1-\delta_1}\,.
    \label{ns0}
	\ea
	After performing algebraic calculations, the above equation can be rewritten as follows
	\ba
	n_s-1 \simeq -6\e_{V} + 2 \eta_{V} + \frac{4}{3} V \xi_{,\phi} \left(\frac{V_{,\phi}}{V}+\frac{2 \xi_{,\phi\phi}}{\xi}\right)\,,
	\label{ns}
	\ea
 in which  $\e_V \equiv \left( {1}/{2}\right) \left( {V_{,\phi}}/{V}\right)^2$ and $\eta_V \equiv {V_{,\phi\phi}}/{V}$ are the potential slow-roll parameters in the standard inflationary model. Note that Eq.~(\ref{ns}) for $\xi=0$  comes back to its counterpart in the standard model.
 The value of the scalar spectral index, based on Planck 2018 TT, TE, EE + lowE + lensing + BK15 + BAO \cite{akrami:2020}, is $n_s=0.9668\pm 0.0037~$ $(68\,\%\,\rm CL)$.\\
Under the slow-roll approximation, the tensor power spectrum ${\cal P}_t$ and the tensor-to-scalar ratio $r$ in the GB model take the following forms \cite{Kawaguchia:2022,Ps&PtHorndeski,Odintsov:2020nsr,Felice:2011,Jiang:2013,Koh:2014}
	\be
	{\cal P}_t=\frac{H^2}{2\pi^2 Q_t c_t^3}\biggr|_{c_t k=a H}\, ,
	\label{powerspectraT}
	\ee
	\ba
	r \equiv \frac{{\cal P}_t}{{\cal P}_s}\simeq 8\left(2\e_1-\delta_1\right)  \label{r0}\,.
	\ea
    After performing algebraic calculations, the above equation can be rewritten as follows	
	\ba
	r \simeq 16\e_{V} + \frac{64}{3} \xi_{,\phi} V\left(\frac{V_{,\phi}}{V}+\frac{2}{3} \xi_{,\phi} V\right) \label{r}\,.
	\ea
The upper limit on the tensor-to-scalar ratio, based on Planck 2018 TT, TE, EE + lowE + lensing + BK15 + BAO \cite{akrami:2020,Ade:2021} is  $r<0.058~$ $(95\%~ \rm CL)$. It is notable that Eq.~(\ref{r}) for $\xi=0$  recovers the result of standard inflationary model.

	\section{Enhancement of the curvature power spectrum}
	\label{sec3}

As mentioned previously,  generation of  PBHs and secondary GWs requires a significant enhancement in the amplitude of the scalar power spectrum at small scales.
This enhancement can be achieved through a short USR phase during the inflation epoch.
In GB inflation, the transition to the USR phase can be accomplished by selecting an appropriate coupling function and adjusting the model parameters.
The model parameters must be adjusted for two aims. Firstly, it is necessary for the model to be consistent with the Planck observational measurements at the pivot scale. Secondly, at small scales, the scalar power spectrum increases by approximately seven orders of magnitude. To achieve both of these goals, we utilize the following coupling function~\cite{Kawai:2021,Kawaguchia:2022,Zhang:2022}
	\be
	\label{eqxi}
	\xi(\phi)=\xi_0\tanh\left[\xi_1(\phi-\phi_c)\right]
	\,,
	\ee
where $\phi_{c}~[\Mpl]$, $\xi_{1}~[\Mpl^{-1}]$ and $\xi_{0}$ are used to characterize the position, width, and height of the formed peak in the scalar power spectrum, respectively.

    In this study, we consider the fractional power-law inflationary potentials as
    \be
    \label{eqv}
    V(\phi)=V_0 \phi^n,
    \ee
where the power index $n$ is set to $\{1/3,~ 2/5,~ 2/3\}$. Here, $V_{0}~[\Mpl^{4-n}]$ can be determined through the constraint of the scalar power spectrum ${\cal P}_s(k_{*}) = 2.1\times{10}^{-9}$ at the pivot scale $k_{*}=0.05~\rm Mpc^{-1}$. It is noteworthy to mention that if the inflaton trajectory passes near a fixed point, it will enter an USR regime. The presence of the GB coupling term, might lead to a non-trivial fixed point.
Around this fixed point, the parameters $\dot{H}$, $\dot{\phi}$, and $\ddot{\phi}$ all become zero. By considering these conditions and using Eq.~(\ref{bge3}), we can deduce the following condition at $\phi=\phi_c$ \cite{Kawai:2021,Kawaguchia:2022,Zhang:2022}

	\be
	\Bigg[ V_{,\phi}+\frac{4 \xi_{,\phi}{V(\phi)}^2}{3\Mpl^4}\Bigg]
	\Biggr|_{\phi=\phi_c}
	=0\,.
	\label{EqC}
	\ee
Utilizing Eq.~(\ref{EqC}), we can specify and adjust the parameters to achieve both aims of this study.
As in the vicinity of a fixed point, when the inflaton undergoes the USR phase  $(\e_2 > 1)$, the slow-roll approximation is not valid, as shown in Figs.~\ref{ETA1Per3}, \ref{ETA2Per5} and \ref{ETA2Per3}. Hence, we cannot utilize Eq.~(\ref{powerspectraS}) to calculate the power spectrum of curvature fluctuations during this era. Hence, we have to solve numerically Eqs.~(\ref{bge2}) and (\ref{bge3}) to estimate the evolution of $H$ and $\phi$ with respect to $e$-fold number $N$ using $dN = Hdt$. The initial conditions for solving background equations are determined by the first Friedmann Eq.~(\ref{bge1}) and slow-roll approximation. We assume that the inflationary period lasts for a duration of 60 $e$-folds from the CMB horizon crossing $N_{*}=0$  to end of inflation $N_{\rm end}$. End of inflation for each case of the model is satisfied by the condition $\e_1 = 1$ as shown in Figs.~\ref{EPSILON1Per3}, \ref{EPSILON2Per5} and \ref{EPSILON2Per3}. Furthermore, to obtain an accurate value for the curvature power spectrum, it is necessary to solve the following Mukhanov-Sasaki (MS) equation numerically
	\be
	u_k''+\left( k^2-\frac{Z_s''}{Z_s}\right)u_k=0\,,
	\label{M.S}
	\ee
	with
	\be
	u_k=Z_s \zeta_k\,,
	\qquad
	Z_s=a \sqrt{2Q_s c_s}\,,
	\label{ZS}
	\ee
in which the prime means the derivative versus the conformal time $\tau_s=\int\left({c_s}/{a}\right){\rm d}t$.
It is important to note that the change in curvature fluctuations  $\zeta_k$, in Fourier space $u_{k}$ during the inflation from sub-horizon scales $c_s k\ll aH$ to super-horizon scales $c_s k\gg aH$ is calculated by using MS equation~(\ref{M.S}).
At the sub-horizon scale, we choose the Fourier transformation of the Bunch-Davies vacuum state as the initial condition as follows
	\be
	u_k=\frac{1}{\sqrt{2 k}}e^{-ik\tau_s}\,.
	\label{BunchDavis}
	\ee
    Consequently, numerical solution of the MS equation (\ref{M.S}) gives rise to the precise value of the scalar power spectrum for each mode $u_k$ as follows
	\be
	{\cal P}_{s}(k)\equiv \frac{k^3}{2\pi^2} \left| \zeta_k (\tau_{s},{ k})\right|^2\,.\label{powerspectraZeta}
	\ee
Using the solutions of the MS equation, in Figs.~\ref{Fig132}, \ref{Fig252} and \ref{Fig232} we plot the scalar power spectrum as a function of $k$ for all cases of the model.

In what follows, we focus on fractional power-law potentials with power indexes $\{1/3$, $2/3$, $2/5\}$.
Finally, we estimate the precise value of the scalar power spectrum by solving the MS equation (\ref{M.S})  and investigate the possibility of PBH formation for the selected  potentials.

	\subsection{$V(\phi)=V_0 \phi^{1/3}$}

As the first class of fractional power-law  potentials, we set $n=1/3$ in Eq.~(\ref{eqv}). Hence, the condition Eq.~(\ref{EqC}) could be rewritten as follows
%
	\be
	\xi_0=\frac{-1}{{4 V_{0}}{\xi_1}{\phi_c}^{4/3}}~.
	\label{EqC13}
	\ee
By employing the above relation, the value of $\xi_0$ can be estimated for arbitrary values of $ \xi_1 $ and $ \phi_c $. However, it is necessary to  fine-tune the estimated value of $\xi_0$ slightly, to achieve an adequate PBH abundance.

In Table~\ref{tab131}, we have listed the adjusted values of $\phi_c$, $\xi_1$, and $\xi_0$ for the case  $n=1/3$. The last column of Table~\ref{tab131} indicates the  value of $\xi_0$ estimated by Eq.~(\ref{EqC13}).
As shown in Table~\ref{tab131}, we have specified three parameter sets for this class of potential, which can explain the PBHs production.
For each set, we have computed the values of the scalar spectral index and the tensor-to-scalar ratio at the pivot scale using Eqs.~(\ref{ns}) and (\ref{r}). Our calculations lead to $n_s=0.971$ and $r=0.033$, which are consistent with the $68\,\%\,\rm CL$ of the Planck  2018  TT, TE, EE + lowE + lensing + BK15 +BAO data \cite{akrami:2020,Ade:2021}.
It is worth noting that,  for all parameter sets of this case of the model, the value of $r$ is also consistent with the recent $95\,\%\,\rm CL$ of the BCEP/Keck 2018 (BK18) constraint $r<0.036$ \cite{Ade:2021}.

\begin{figure}[H]
	\begin{minipage}[b]{1\textwidth}
		\subfigure[\label{PHI1Per3} ]{\includegraphics[width=0.45\textwidth]%
			{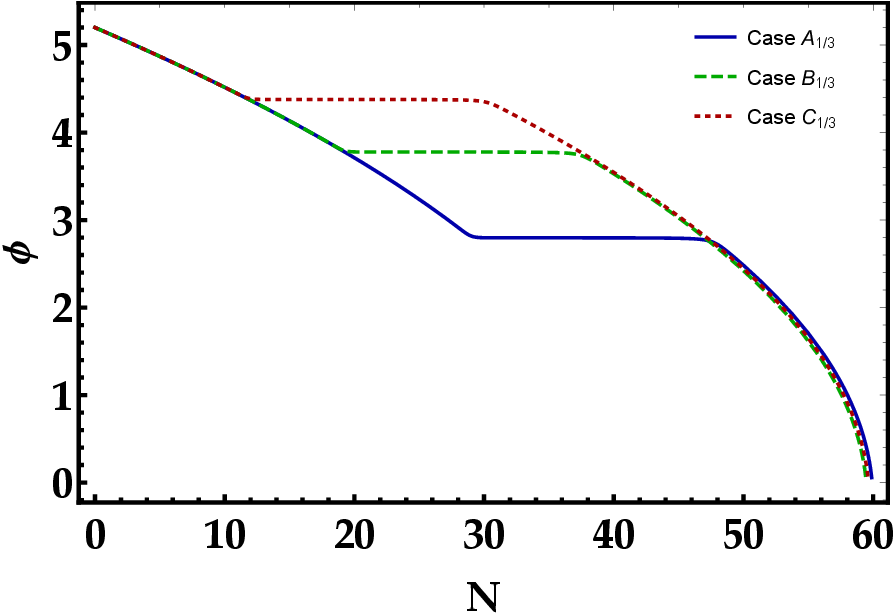}}
		\hspace{.1cm}
		\subfigure[\label{H1Per3}]{\includegraphics[width=.49\textwidth]%
			{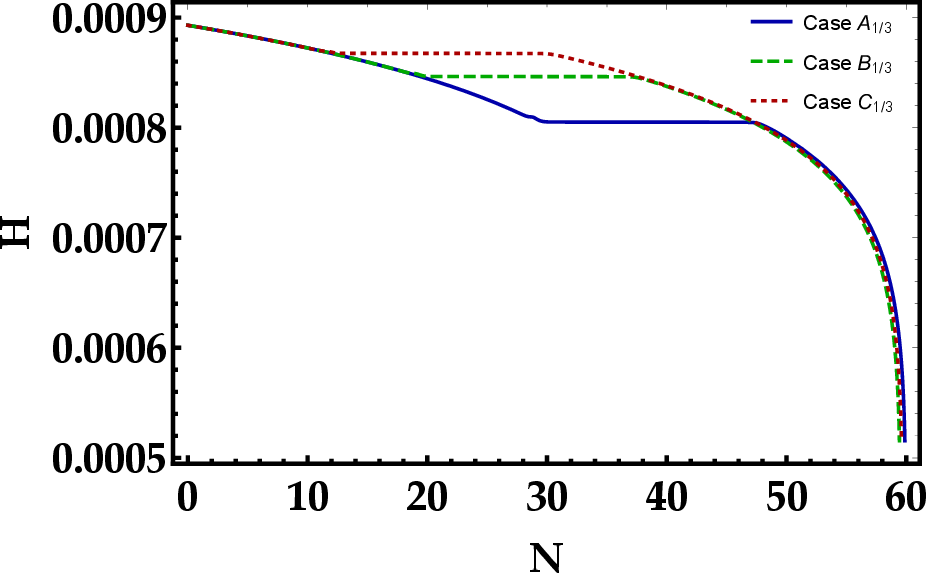}}
		\subfigure[\label{ETA1Per3}]{\includegraphics[width=.47\textwidth]%
			{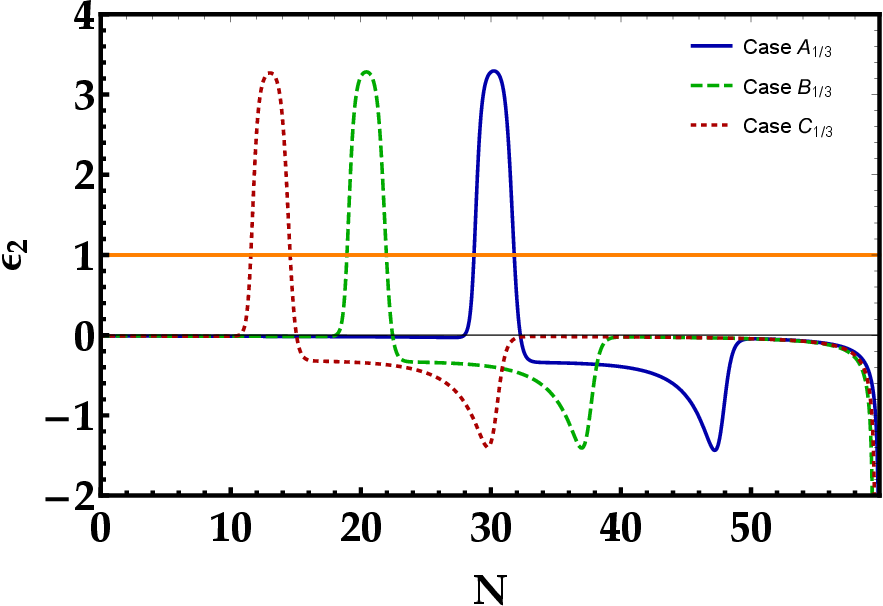}}
		\hspace{.6cm}
		\subfigure[\label{EPSILON1Per3}]{\includegraphics[width=.48\textwidth]%
			{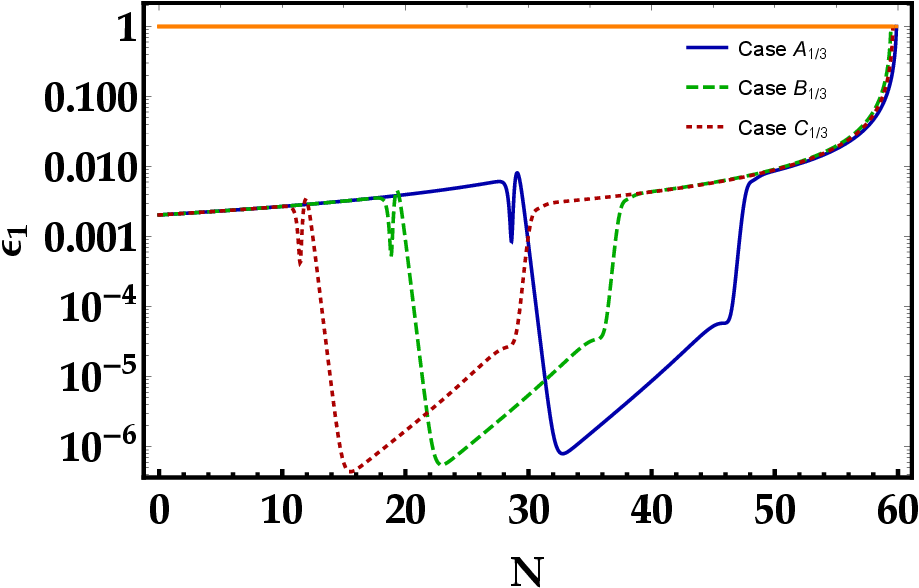}}
		\subfigure[\label{CS1Per3}]{\includegraphics[width=.47\textwidth]%
			{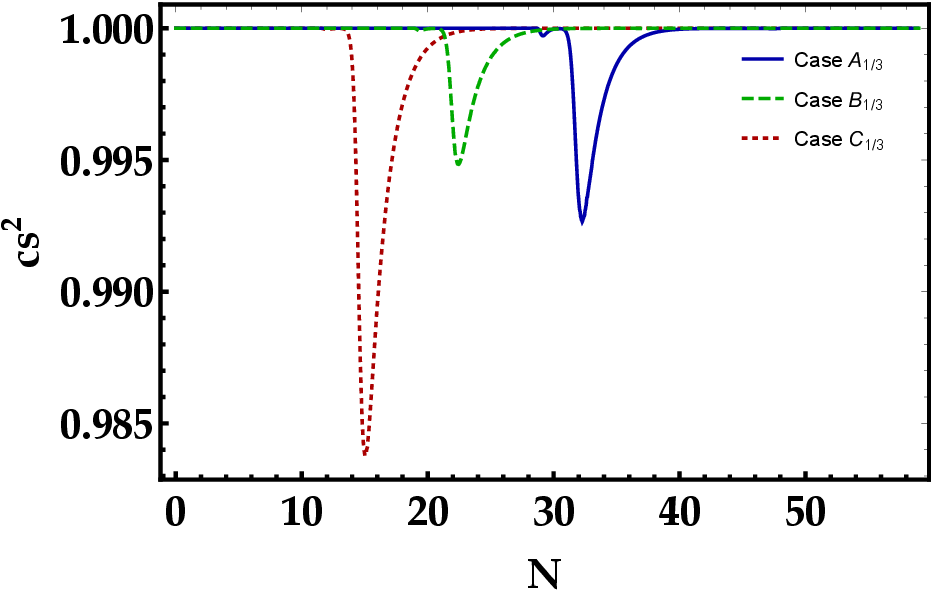}}
		\hspace{.6cm}
		\subfigure[\label{Ct1Per3}]{\includegraphics[width=.48\textwidth]%
			{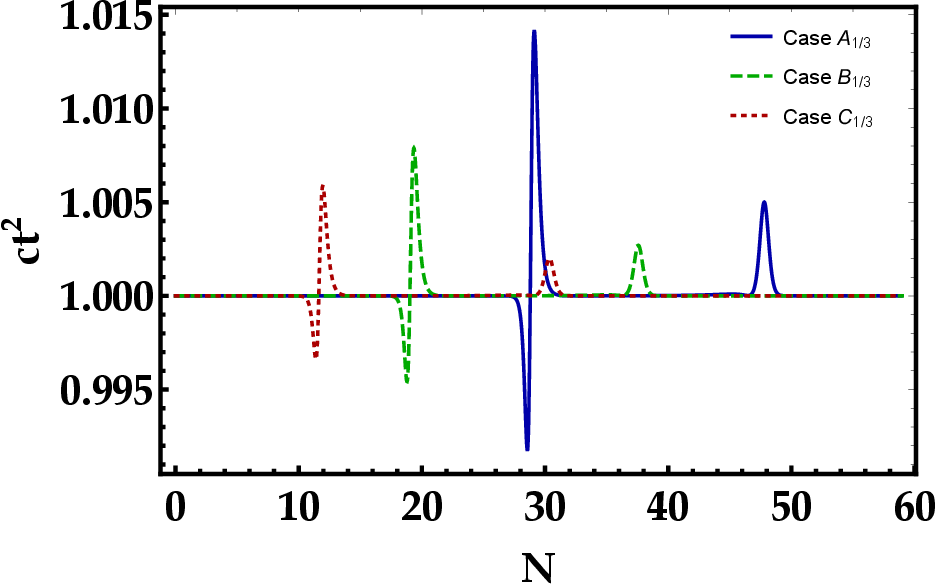}}
		
	\end{minipage}
		\caption{Evolution of (a) the scalar filed $\phi$, (b) the Hubble parameter, (c) the second slow-roll parameter $\e_2$, (d) the first slow-roll parameter $\e_1$, (e) $c_s^{2}$, (f) $c_t^{2}$  versus the $e$-fold number $N$ for three parameter sets of Table~\ref{tab131} for the case $n=1/3$. The blue, green and red lines are corresponding to  $\rm{Case~}A_{\rm 1/3}$, $\rm{Case~}B_{\rm 1/3}$ and $\rm{Case~}C_{\rm 1/3}$, respectively.}
		\label{Fig131}
	\end{figure}

Additionally, the evolution of the scalar field $\phi$ and Hubble parameter $H$ is shown in Figs. \ref{PHI1Per3} and \ref{H1Per3}, respectively. The flat region in these figures describes the USR phase, where the velocity of inflation decreases and the primordial curvature perturbations significantly increase. As mentioned before, this process leads to the violation of the slow-roll condition, which is obvious from Fig. \ref{ETA1Per3}.
Indeed, although the value of $ \e_1 $ remain less than one (see Fig. \ref{EPSILON1Per3}), the value of $\e_2$ increases and exceeds one, which indicates the violation of the slow-roll condition.
Furthermore, we plot the evolution of  $c_s^{2}$ and  $c_t^{2}$ in Figs. \ref{CS1Per3} and \ref{Ct1Per3}, respectively. These figures demonstrate that our model is consistent with stability conditions in Eq.~(\ref{avoidinstability}).

\begin{figure}[H]
		\centering
		\hspace*{-1cm}
		\includegraphics[scale=0.55]{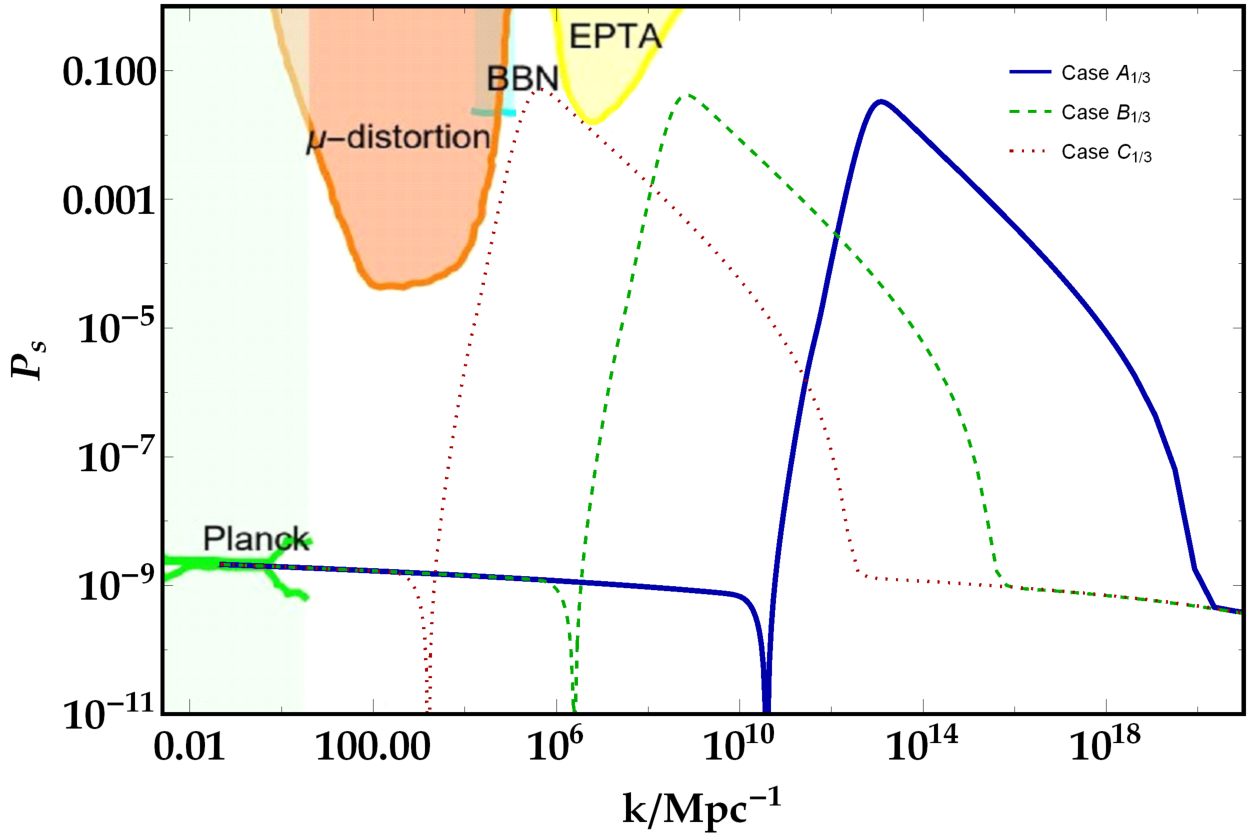}
		\caption{The scalar power spectra ${\cal P}_{s}$ in terms of wavenumber $k$, for parameter sets  of case $n=1/3$ listed in Table~\ref{tab131}.  The blue, green and red lines are corresponding to $\rm{Case~}A_{\rm 1/3}$, $\rm{Case~}B_{\rm 1/3}$ and $\rm{Case~}C_{\rm 1/3}$, respectively. The CMB observations foreclose the light-green shaded region \cite{akrami:2020}. The orange region shows the $\mu$-distortion of CMB \cite{Fixsen:1996}. The cyan area represents the effect on the ratio between neutron and proton during the big bang nucleosynthesis (BBN) \cite{Inomata:2016}. The EPTA observations constrain the yellow area \cite{Inomata:2019-a}. }
		\label{Fig132}
	\end{figure}

Finally, after solving MS equation~(\ref{M.S}),  the value of the scalar power spectrum at the peak position ${\cal P}_{s}^\text{peak}$ is estimated and listed in Table \ref{tab132}.
In addition, in Fig.~\ref{Fig132} the evolution of ${\cal P}_{s}$ is plotted for all parameter sets listed in Table \ref{tab131}. The results indicate that the scalar power spectrum is compatible with CMB measurements at the pivot scale. Moreover, at small scales  ${\cal P}_{s}$ enhances by around seven orders of magnitude, which is completely suitable for PBH formation.	
	
		\begin{table}[H]
		\centering
		\caption{The parameter sets for the case $n=1/3$. For all sets of this case  $V_0=1.38\times 10^{-6}$ $~\Mpl^{11/3}$, $\phi_{*} = 5.2\Mpl$, $n_s=0.971$ and $r=0.033$. The values of $\phi_{*} $, $n_s$ and $r$ are obtained at the CMB horizon exit $N_{*} = 0$.}
		\begin{tabular}{ccccc}
			\hline
			Sets \quad &\quad $\phi_{c}$$\left[ \Mpl\right]$ \quad & \quad $\xi_{1}$$\left[\Mpl^{-1}\right]$\quad &$\xi_{0}$\quad & $\xi_{0} \left(\rm From~Eq.~(\ref{EqC13})\right)$\quad  \\ [0.5ex]
			\hline
			\hline
			$\rm{Case~}A_{\rm 1/3}$ \quad &\quad $2.80$ \quad &\quad $23.0$ \quad &\quad $-2.0052\times10^{3}$&\quad $-1.9958\times10^{3}$ \quad\\[0.5ex]
			\hline
			$\rm{Case~}B_{\rm 1/3}$ \quad &\quad $3.78$ \quad &\quad $30.5$\quad &\quad $-1.0133\times10^{3}$&\quad $-1.0087\times10^{3}$ \quad\\[0.5ex]
			\hline
			$\rm{Case~}C_{\rm 1/3}$ \quad &\quad $4.38$ \quad &\quad $35.0$\quad &\quad $-7.2538\times10^{2}$&\quad $-7.2226\times10^{2}$ \quad\\[0.5ex]
			\hline
			
		\end{tabular}
		\label{tab131}
	\end{table}

	\begin{table}[H]
		\centering
		\caption{The values of $k_{\text{peak}}$, ${\cal P}_{s}^\text{peak}$, $M_{\text{PBH}}^{\text{peak}}$ and $f_{\text{PBH}}^{\text{peak}}$ for the cases listed in Table \ref{tab131}.}
		\begin{tabular}{ccccccc}
			\hline
			Sets \quad & \quad$k_{\text{peak}}/\text{\rm Mpc}^{-1}$ \quad &\quad ${\cal P}_{s}^\text{peak}$ \quad & \quad  $M_{\text{PBH}}^{\text{peak}}/M_{\odot}$ \quad & \quad $f_{\text{PBH}}^{\text{peak}}$\\ [0.5ex]
			\hline
			\hline
			$\rm{Case~}A_{\rm 1/3}$ \qquad &\quad $1.1854\times10^{13}$ \quad &\quad $0.0328$ \quad &\quad $1.93\times 10^{-14}$ \quad &$0.9950$ \\[0.5ex]
			\hline
			$\rm{Case~}B_{\rm 1/3}$ \quad &\quad $6.9008\times10^{8}$ \quad &\quad $0.0417$ \quad &\quad $5.94\times10^{-6}$ \quad & $0.0539$ \\[0.5ex]
			\hline
			$\rm{Case~}C_{\rm 1/3}$ \quad &\quad $4.3389\times10^{5}$ \quad &\quad $0.0514$ \quad &\quad $15.80$ \quad &$0.0017$\\[0.5ex]
			\hline
		\end{tabular}
		\label{tab132}
	\end{table}

    \subsection{$V(\phi)=V_0 \phi^{2/5}$}
    \space
  Here, by setting $n=2/5$ in Eq.~(\ref{eqv}), the condition (\ref{EqC}) is reformed as follows

	\be
	\xi_0=\frac{-3}{{10 V_{0}}{\xi_1}{\phi_c}^{7/5}}.
	\label{EqC25}
	\ee
Similar to the previous case, we utilize the condition given by Eq.~(\ref{EqC25}) to estimate the value of $\xi_0$ for arbitrary values of $\xi_1$ and $\phi_c$. With a slight fine-tuning, the proper value of $\xi_0$ can be achieved.
The fourth column of Table \ref{tab251} displays the calculated values of $\xi_0$ using Eq.~(\ref{EqC25}).
Three parameter sets for this class of our model have been listed in Table \ref{tab251}.
In this class, the estimations indicate that $n_s=0.971$ and
$r=0.038$, which are in agreement with the $68\,\%\,\rm CL$ of the Planck 2018 TT, TE, EE + lowE + lensing + BK15 +BAO data \cite{akrami:2020,Ade:2021}.

\begin{figure}[H]
		\begin{minipage}[b]{1\textwidth}
			\subfigure[\label{PHI2Per5} ]{\includegraphics[width=0.45\textwidth]%
				{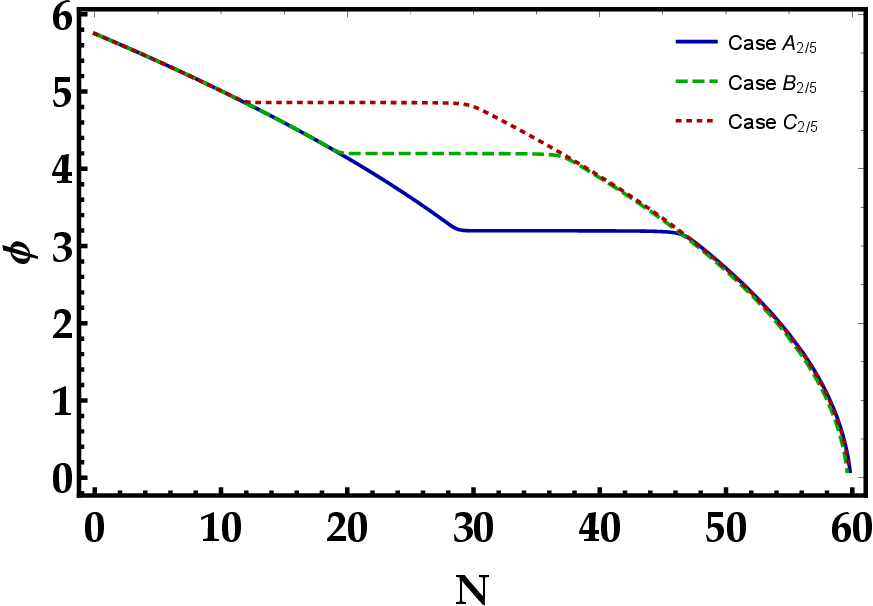}}
			\hspace{.1cm}
			\subfigure[\label{H2Per5}]{\includegraphics[width=.49\textwidth]%
				{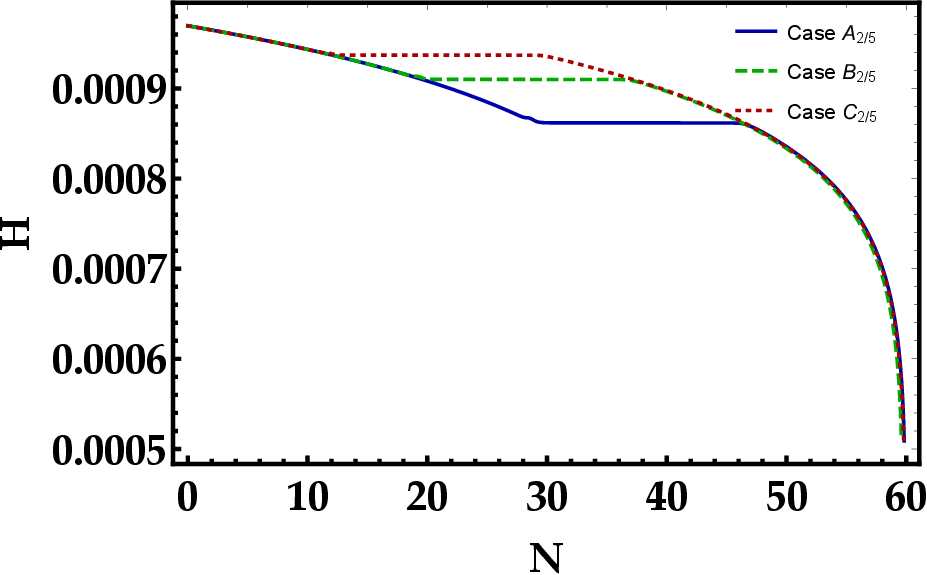}}
			\subfigure[\label{ETA2Per5}]{\includegraphics[width=.47\textwidth]%
				{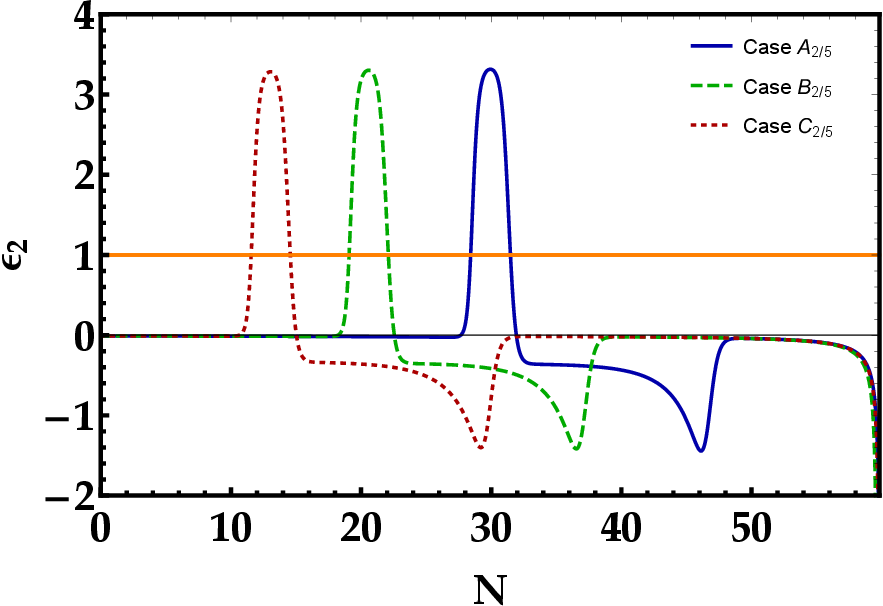}}
			\hspace{.6cm}
			\subfigure[\label{EPSILON2Per5}]{\includegraphics[width=.48\textwidth]%
				{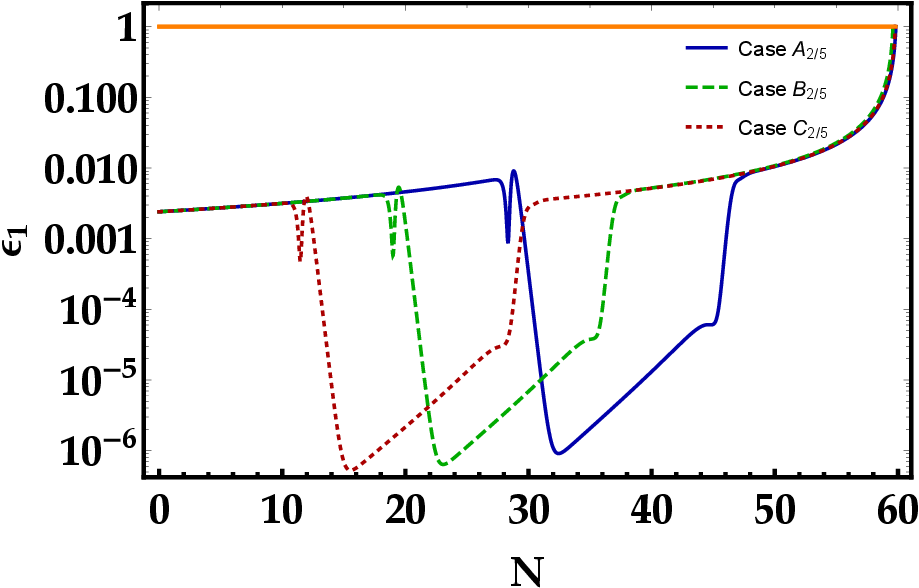}}
			\subfigure[\label{CS2Per5}]{\includegraphics[width=.47\textwidth]%
				{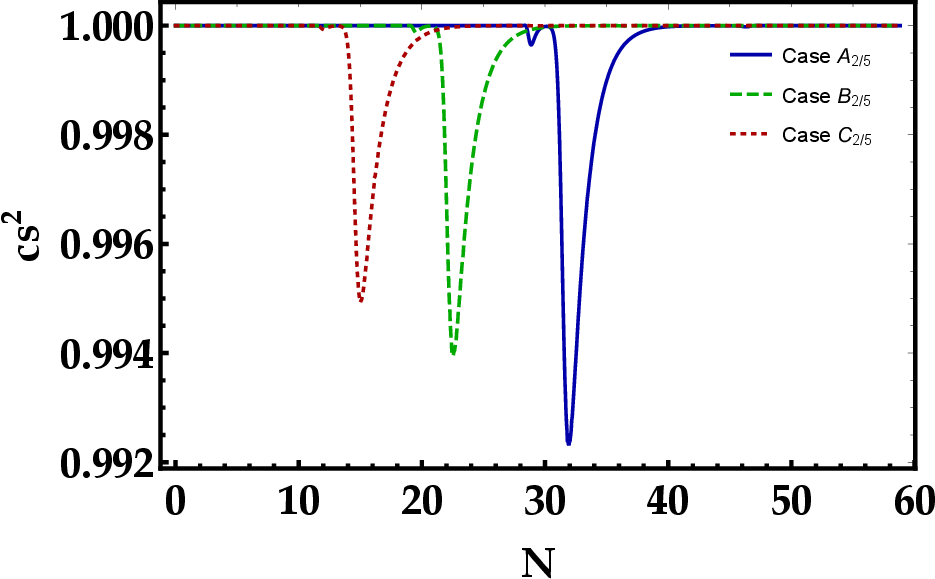}}
			\hspace{.6cm}
			\subfigure[\label{Ct2Per5}]{\includegraphics[width=.48\textwidth]%
				{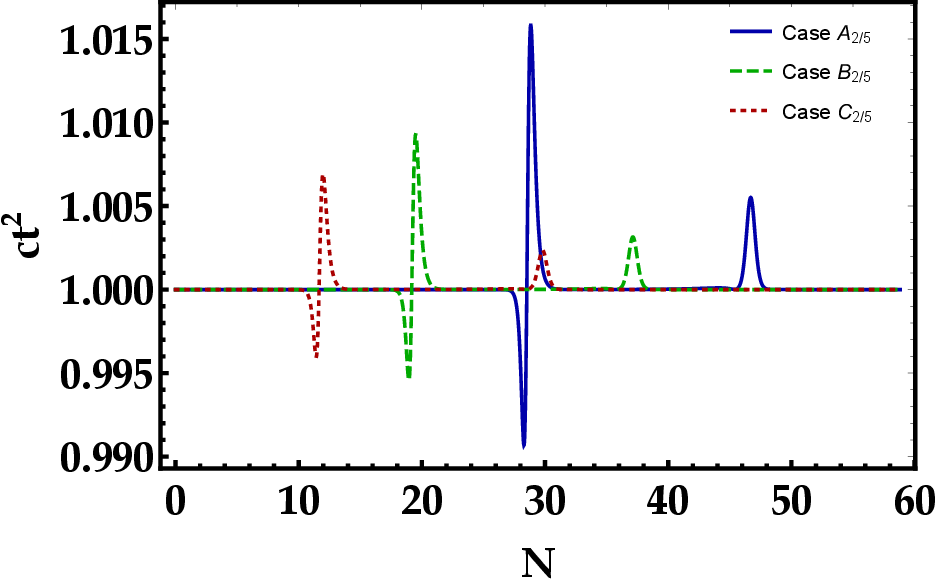}}
			
		\end{minipage}
		\caption{Same as Fig. \ref{Fig131}, but for the case $n=2/5$. The blue, green and red lines are corresponding to  $\rm{Case~}A_{\rm 2/5}$, $\rm{Case~}B_{\rm 2/5}$ and $\rm{Case~}C_{\rm 2/5}$, respectively.
		}\label{Fig251}
	\end{figure}

Figures \ref{PHI2Per5} and \ref{H2Per5} describe the evolution of the scalar field $\phi$ and Hubble parameter $H$, respectively. The flat era demonstrates where the inflaton undergoes the USR regime.
The  behaviours of the slow-roll parameters are displayed in Figs. \ref{ETA2Per5} and \ref{EPSILON2Per5}, where the violation of the slow-roll condition is evident from Fig. \ref{ETA2Per5}. On the other hand, the evolutions of  $c_s^{2}$ and  $c_t^{2}$ are shown in Figs. \ref{CS2Per5} and \ref{Ct2Per5}, respectively. These figures demonstrate good agreement with conditions mentioned in Eq.~(\ref{avoidinstability}).

\begin{figure}[H]
		\centering
		\hspace*{-1cm}
		\includegraphics[scale=0.55]{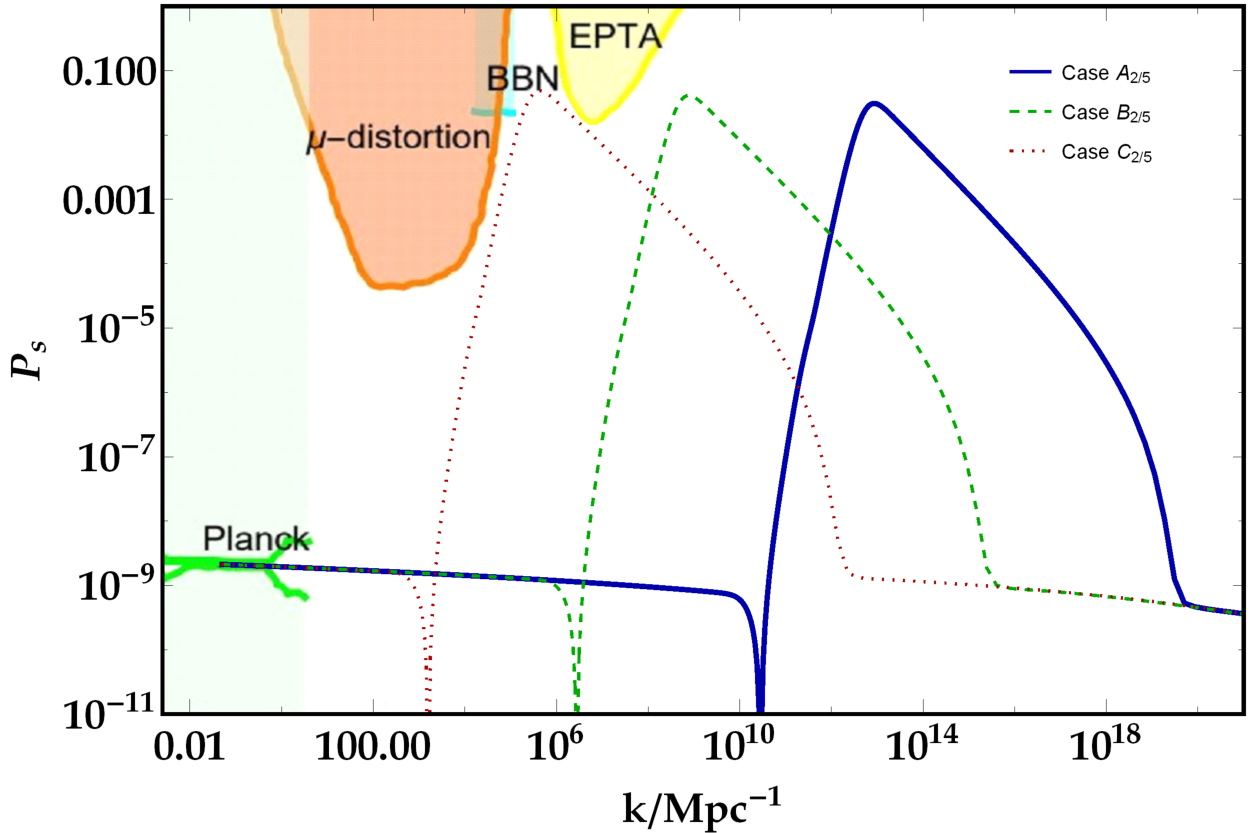}
		\caption{Same as Fig. \ref{Fig132}, but for the case $n=2/5$. The blue, green and red lines  are corresponding to  $\rm{Case~}A_{\rm 2/5}$, $\rm{Case~}B_{\rm 2/5}$ and $\rm{Case~}C_{\rm 2/5}$, respectively.}
		\label{Fig252}
	\end{figure}

Additionally, the values of the scalar power spectra at the peak position ${\cal P}_{s}^\text{peak}$ for all cases listed in Table. \ref{tab251} are shown in Table. \ref{tab252}.
Figure. \ref{Fig252} exhibits the evolution of the ${\cal P}_{s}$ in term of comoving wavenumber $k$.
It is clear that for all cases, the scalar power spectrum has a good agreement with CMB measurements, and its enhancement at small scales is adequate for PBH production.
	
	\begin{table}[H]
		\centering
		\caption{The parameter sets for case $n=2/5$. For all sets of this case  $V_0=1.4\times 10^{-6}$ $~\Mpl^{18/5}$, $\phi_{*} = 5.75\Mpl$, $n_s=0.971$ and
			$r=0.038$. The values of $\phi_{*} $, $n_s$ and $r$ are obtained at the CMB horizon exit $N_{*} = 0$.}
		\begin{tabular}{ccccc}
			\hline
			Sets \quad &\quad $\phi_{c}$$\left[ \Mpl\right]$ \quad & \quad $\xi_{1}$$\left[\Mpl^{-1}\right]$\quad &$\xi_{0}$\quad & $\xi_{0} \left(\rm From~Eq.~(\ref{EqC25})\right)$\quad  \\ [0.5ex]
			\hline
			\hline
			$\rm{Case~}A_{\rm 2/5}$ \quad &\quad $3.20$ \quad &\quad $22.1$ \quad &\quad $-1.9130\times10^{3}$&\quad $-1.9027\times10^{3}$ \quad\\[0.5ex]
			\hline
			$\rm{Case~}B_{\rm 2/5}$ \quad &\quad $4.20$ \quad &\quad $28.5$ \quad &\quad $-1.0135\times10^{3}$&\quad $-1.0083\times10^{3}$ \quad\\[0.5ex]
			\hline
			$\rm{Case~}C_{\rm 2/5}$ \quad &\quad $4.86$ \quad &\quad $32.6$\quad &\quad $-7.2201\times 10^{2}$&\quad $-7.1859\times10^{2}$ \quad\\[0.5ex]
			\hline
		\end{tabular}
		\label{tab251}
	\end{table}

	\begin{table}[H]
		\centering
		\caption{The values of $k_{\text{peak}}$, ${\cal P}_{s}^\text{peak}$, $M_{\text{PBH}}^{\text{peak}}$ and $f_{\text{PBH}}^{\text{peak}}$ for the cases listed in Table \ref{tab251}.}
		\begin{tabular}{ccccccc}
			\hline
			Sets \quad & \quad$k_{\text{peak}}/\text{\rm Mpc}^{-1}$ \quad & \quad ${\cal P}_{s}^\text{peak}$\quad &\quad $M_{\text{PBH}}^{\text{peak}}/M_{\odot}$ \quad  & \quad $f_{\text{PBH}}^{\text{peak}}$\\ [0.5ex]
			\hline
			\hline
			$\rm{Case~}A_{\rm 2/5}$ \qquad &\quad $8.6519\times10^{12}$ \quad &\quad $0.0309$ \quad &\quad  $4.25\times 10^{-14}$ \quad &$0.9950$ \\[0.5ex]
			\hline
			$\rm{Case~}B_{\rm 2/5}$ \quad &\quad $7.5544\times10^{8}$ \quad &\quad $0.0412$ \quad &\quad $4.85\times10^{-6}$ \quad & $0.0644$ \\[0.5ex]
			\hline
			$\rm{Case~}C_{\rm 2/5}$ \quad &\quad $4.3008\times10^{5}$ \quad &\quad $0.0486$ \quad &\quad $15.84$ \quad &$0.0019$\\[0.5ex]
			\hline
		\end{tabular}
		\label{tab252}
	\end{table}

	\newpage
	\subsection{$V(\phi)=V_0 \phi^{2/3}$}
	\space
Here, we set $n=2/3$, and consequently, the condition (\ref{EqC}) can be rewritten as follows
	\be
	\xi_0=\frac{-1}{{2 V_{0}}{\xi_1}{\phi_c}^{5/3}}~.
	\label{EqC23}
	\ee
In Table~\ref{tab231}, we show the adjusted parameters $\phi_{c}$, $\xi_{1}$, and $\xi_{0}$ for all cases of class $n=2/3$.
The last column in Table \ref{tab231} shows the value of $\xi_{0}$ computed with Eq.~(\ref{EqC23}).
We have adjusted the value of $\xi_{0}$ to be suitable for the generation of PBHs, and the results are depicted in the third column of Table \ref{tab231}.
In this class , the estimations indicate that $n_s=0.971$ and
$r=0.056$ for all parameter sets. These values are respectively in agreement with the $95\,\%\,\rm CL$ of the Planck 2018 TT, TE, EE + lowE + lensing + BK15 +BAO data \cite{akrami:2020,Ade:2021}.

\begin{figure}[H]
		\begin{minipage}[b]{1\textwidth}
			\subfigure[\label{PHI2Per3} ]{\includegraphics[width=0.45\textwidth]%
				{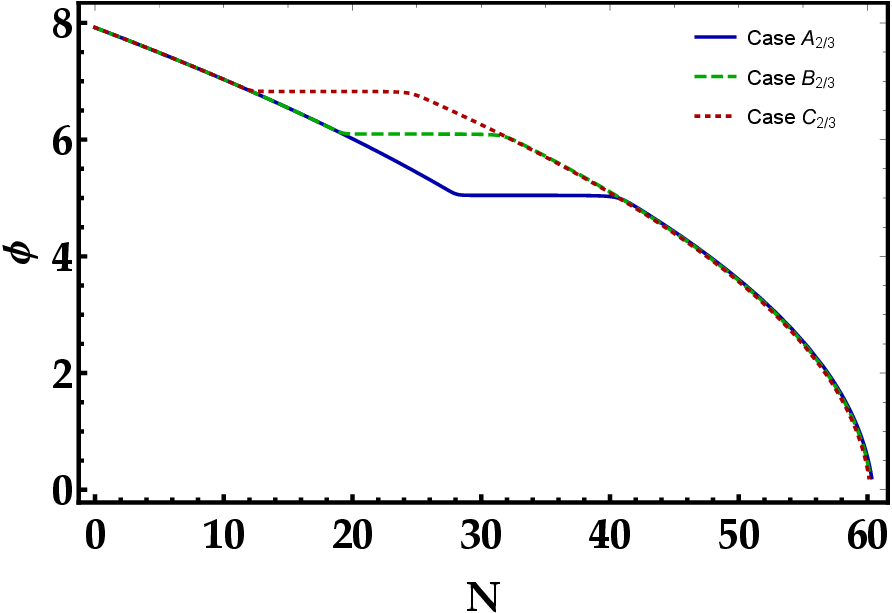}}
			\hspace{.1cm}
			\subfigure[\label{H2Per3}]{\includegraphics[width=.49\textwidth]%
				{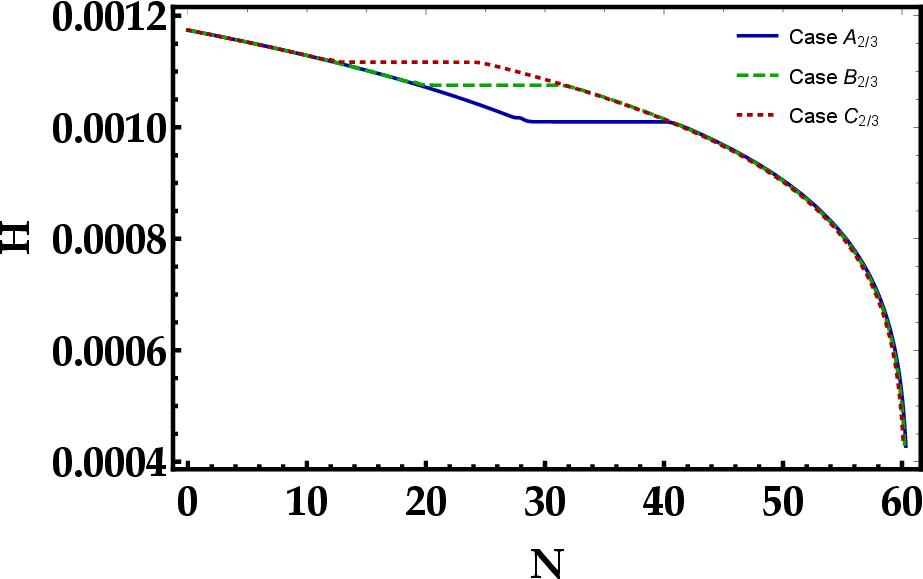}}
			\subfigure[\label{ETA2Per3}]{\includegraphics[width=.47\textwidth]%
				{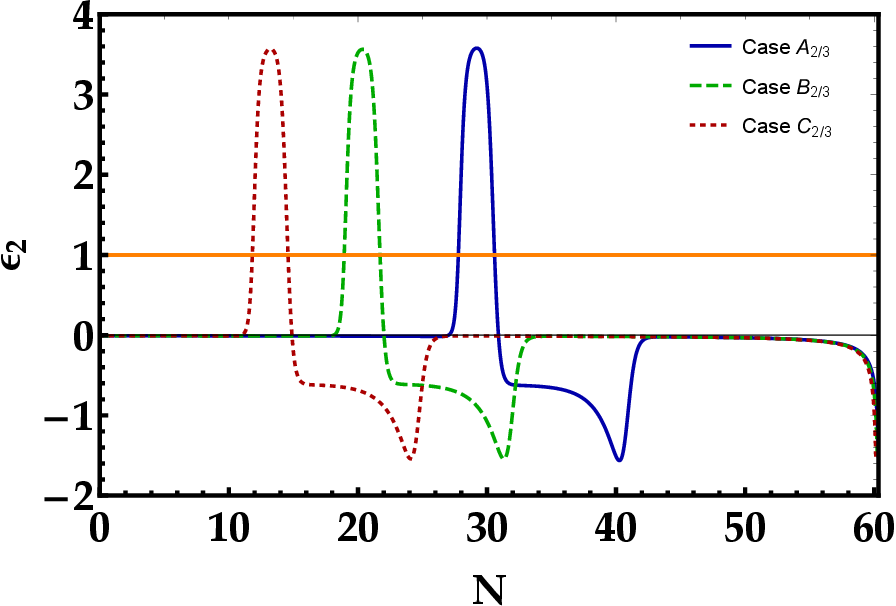}}
			\hspace{.6cm}
			\subfigure[\label{EPSILON2Per3}]{\includegraphics[width=.48\textwidth]%
				{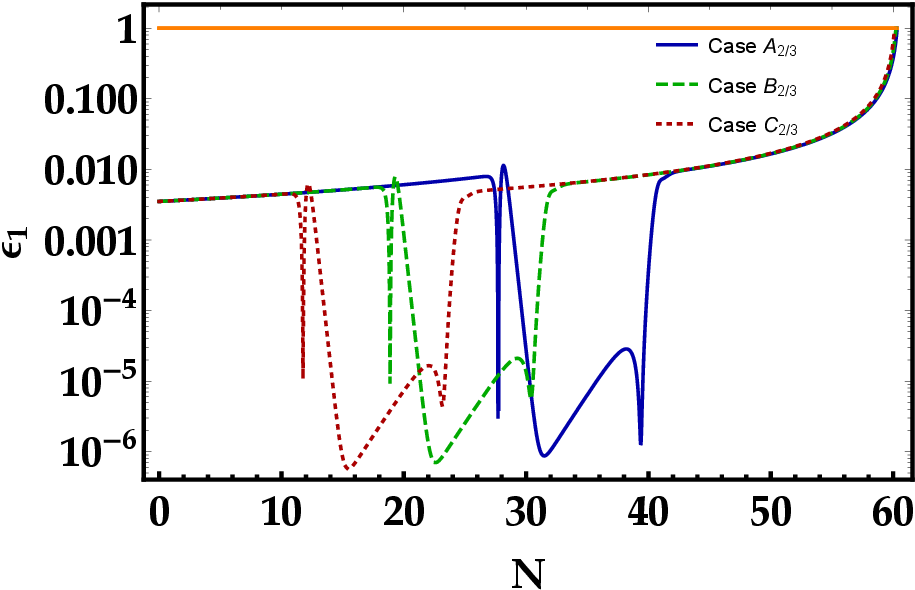}}
			\subfigure[\label{Cs2Per3}]{\includegraphics[width=.47\textwidth]%
				{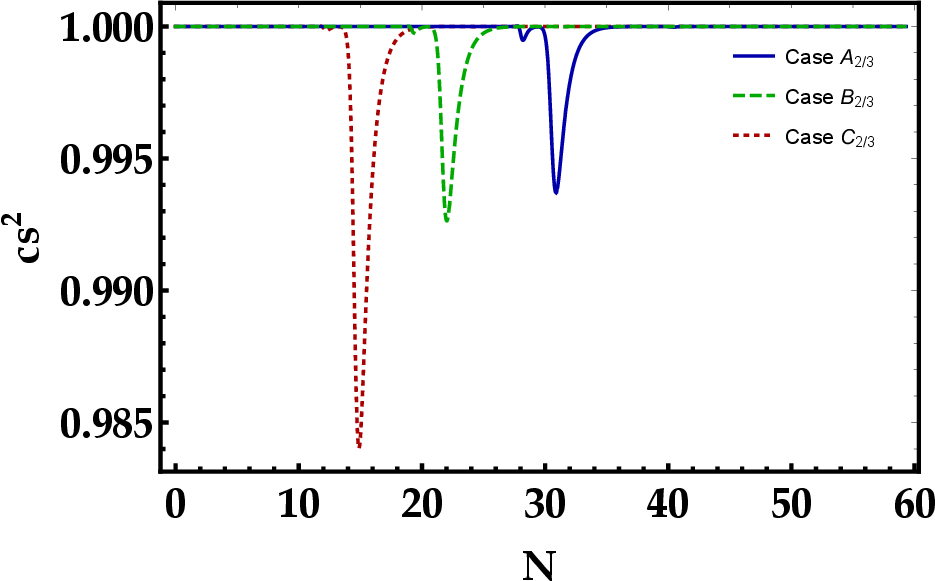}}
			\hspace{.6cm}
			\subfigure[\label{Ct2Per3}]{\includegraphics[width=.48\textwidth]%
				{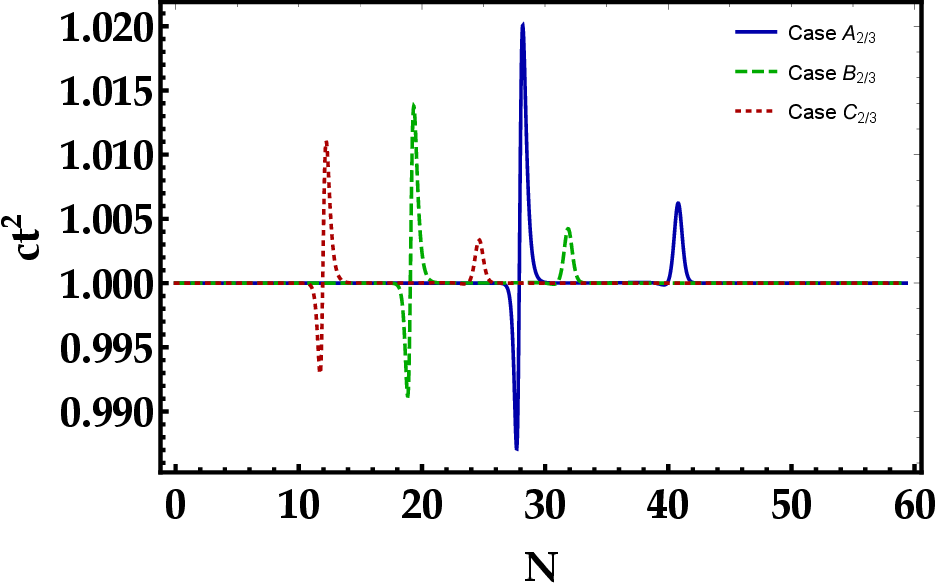}}
			
		\end{minipage}
		\caption{Same as Fig. \ref{Fig131}, but for the case $n=2/3$. The blue, green and red lines are corresponding to  $\rm{Case~}A_{\rm 2/3}$, $\rm{Case~}B_{\rm 2/3}$ and $\rm{Case~}C_{\rm 2/3}$, respectively. }
		\label{Fig231}
	    \end{figure}

\begin{figure}[H]
		\centering
		\hspace*{-1cm}
		\includegraphics[scale=0.55]{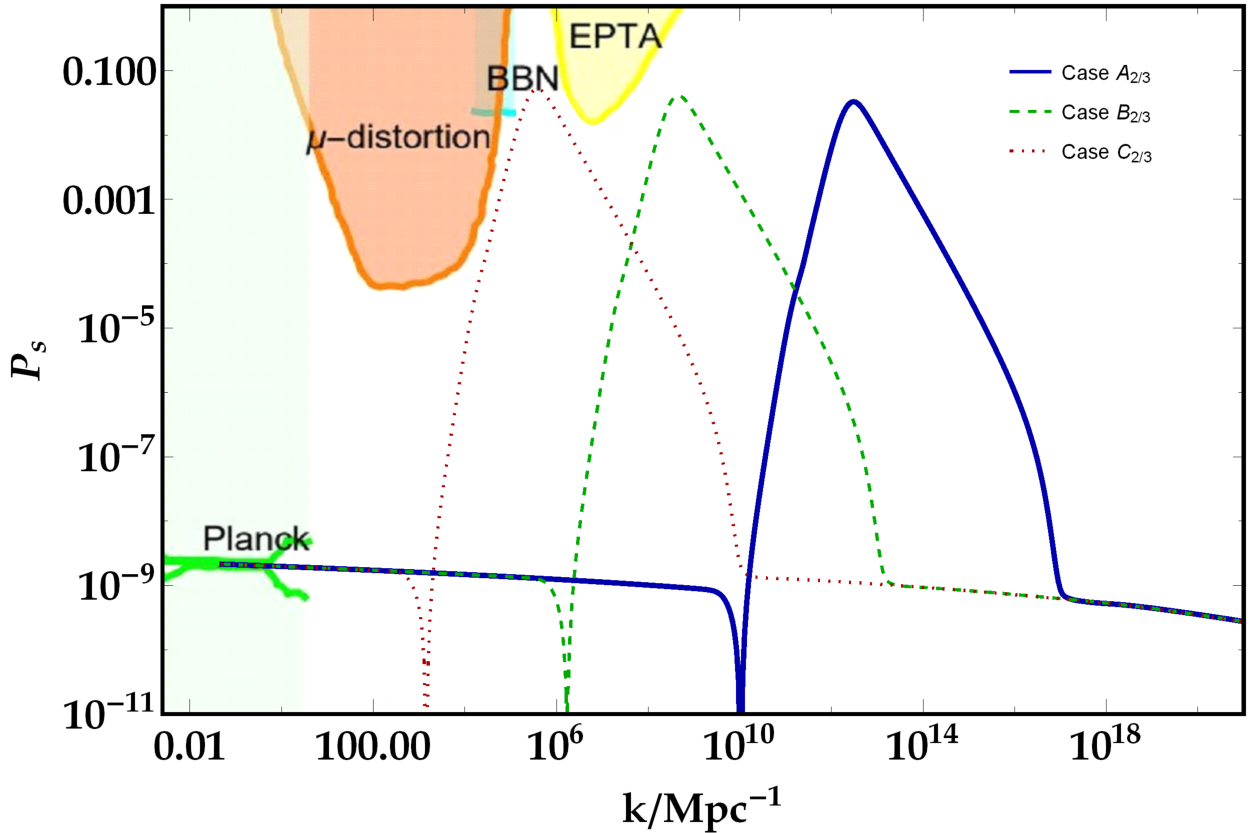}
		\caption{Same as Fig. \ref{Fig132}, but for the case $n=2/3$. The blue, green and red lines are corresponding to  $\rm{Case~}A_{\rm 2/3}$, $\rm{Case~}B_{\rm 2/3}$ and $\rm{Case~}C_{\rm 2/3}$, respectively.}
		\label{Fig232}
	\end{figure}

We plot the evolutions of the scalar field and Hubble parameter in Figs. \ref{PHI2Per3} and \ref{H2Per3}, respectively.
The first and second slow-roll parameters are shown in Figs. \ref{EPSILON2Per3} and \ref{ETA2Per3}.  As shown in Fig. \ref{ETA2Per3}, in a short region $\e_2 $ becomes larger than one, which means that the slow-roll condition is violated.
In Figs. \ref{Cs2Per3} and \ref{Ct2Per3}, we investigate the behaviour of the $c_s^2$ and $c_t^2$ to check the ghosts and Laplacian instabilities. The results confirms consistency of our model with condition given by Eq.~(\ref{avoidinstability}).
In Table \ref{tab232}, the exact values of power spectra at the peak position are listed for all cases in Table \ref{tab231}.
Additionally, we plot ${\cal P}_s$ in terms of comoving wavenumber for all cases of this class in Fig. \ref{Fig232}.
	
	\begin{table}[H]
		\centering
		\caption{The parameter sets for the case $n=2/3$. For all sets of this case $V_0=1.04\times10^{-6}$ $~\Mpl^{10/3}$, $\phi_{*} = 7.92\Mpl$, $n_s=0.971$ and $r=0.056$. The values of $\phi_{*} $, $n_s$ and $r$ are obtained at the CMB horizon exit $N_{*} = 0$.}
		\begin{tabular}{ccccc}
			\hline
		Sets \quad &\quad $\phi_{c}$$\left[ \Mpl\right]$ \quad & \quad $\xi_{1}$$\left[\Mpl^{-1}\right]$\quad &$\xi_{0}$\quad & $\xi_{0} \left(\rm From~Eq.~(\ref{EqC23})\right)$\quad  \\[0.5ex]
			\hline
			\hline
			$\rm{Case~}A_{\rm 2/3}$ \quad &\quad $5.05$ \quad &\quad $23.20$ \quad &\quad $-1.4161\times10^{3}$&\quad $-1.3941\times10^{3}$ \quad\\[0.5ex]
			\hline
			$\rm{Case~}B_{\rm 2/3}$ \quad &\quad $6.10$ \quad &\quad $27.65$ \quad &\quad $-8.6684\times10^{2}$&\quad $-8.5380\times10^{2}$ \quad\\[0.5ex]
			\hline
			$\rm{Case~}C_{\rm 2/3}$ \quad &\quad $6.83$ \quad &\quad $30.90$ \quad &\quad $-6.4263\times10^{2}$&\quad $-6.3281\times10^{2}$ \quad\\[0.5ex]
			\hline
		\end{tabular}
		\label{tab231}
	\end{table}

	\begin{table}[H]
		\centering
		\caption{The values of $k_{\text{peak}}$, ${\cal P}_{s}^\text{peak}$, $M_{\text{PBH}}^{\text{peak}}$ and $f_{\text{PBH}}^{\text{peak}}$ for the cases listed in Table \ref{tab231}.}
		\begin{tabular}{ccccccc}
			\hline
			Sets \quad & \quad$k_{\text{peak}}/\text{\rm Mpc}^{-1}$ \quad & \quad ${\cal P}_{s}^\text{peak}$\quad &\quad $M_{\text{PBH}}^{\text{peak}}/M_{\odot}$ \quad  & \quad $f_{\text{PBH}}^{\text{peak}}$\\ [0.5ex]
			\hline
			\hline
			$\rm{Case~}A_{\rm 2/3}$ \qquad &\quad $3.0757\times10^{12}$ \quad &\quad $0.0328$ \quad &\quad  $3.26\times 10^{-13}$ \quad &$0.9940$ \\[0.5ex]
			\hline
			$\rm{Case~}B_{\rm 2/3}$ \quad &\quad $4.4693\times10^{8}$ \quad &\quad $0.0408$ \quad &\quad $1.58\times10^{-5}$ \quad & $0.0346$ \\[0.5ex]
			\hline
			$\rm{Case~}C_{\rm 2/3}$ \quad &\quad $3.8356\times10^{5}$ \quad &\quad $0.0519$ \quad &\quad $23.50$ \quad &$0.0015$\\[0.5ex]
			\hline
		\end{tabular}
		\label{tab232}
	\end{table}

	\section{The abundance of PBHs}
	\label{sec4}
\begin{figure}[H]
		\begin{minipage}[b]{1\textwidth}
			\centering
			\subfigure[\label{FPBHs1Per3}]{\includegraphics[width=.49\textwidth]%
				{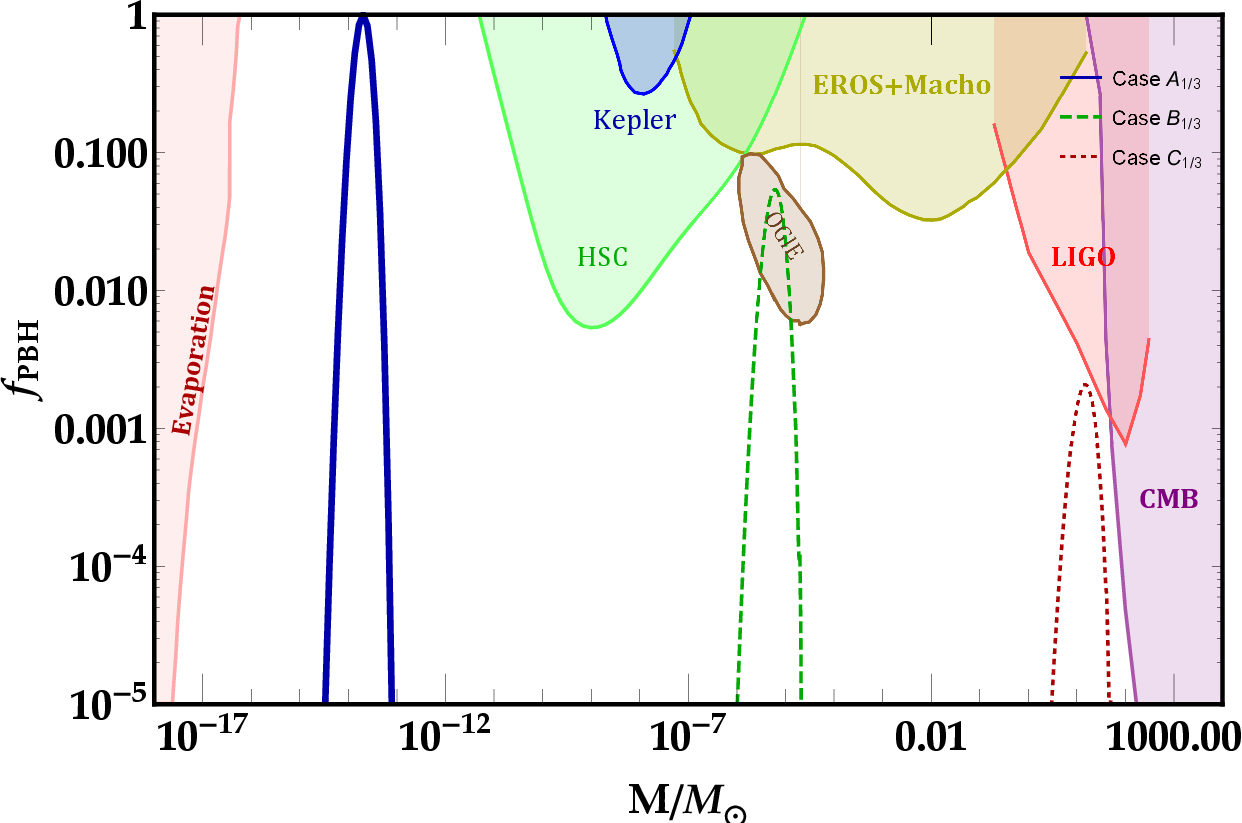}}
			\hspace{.1cm}	
			\centering
			\subfigure[\label{FPBHs2Per5}]{\includegraphics[width=.49\textwidth]%
				{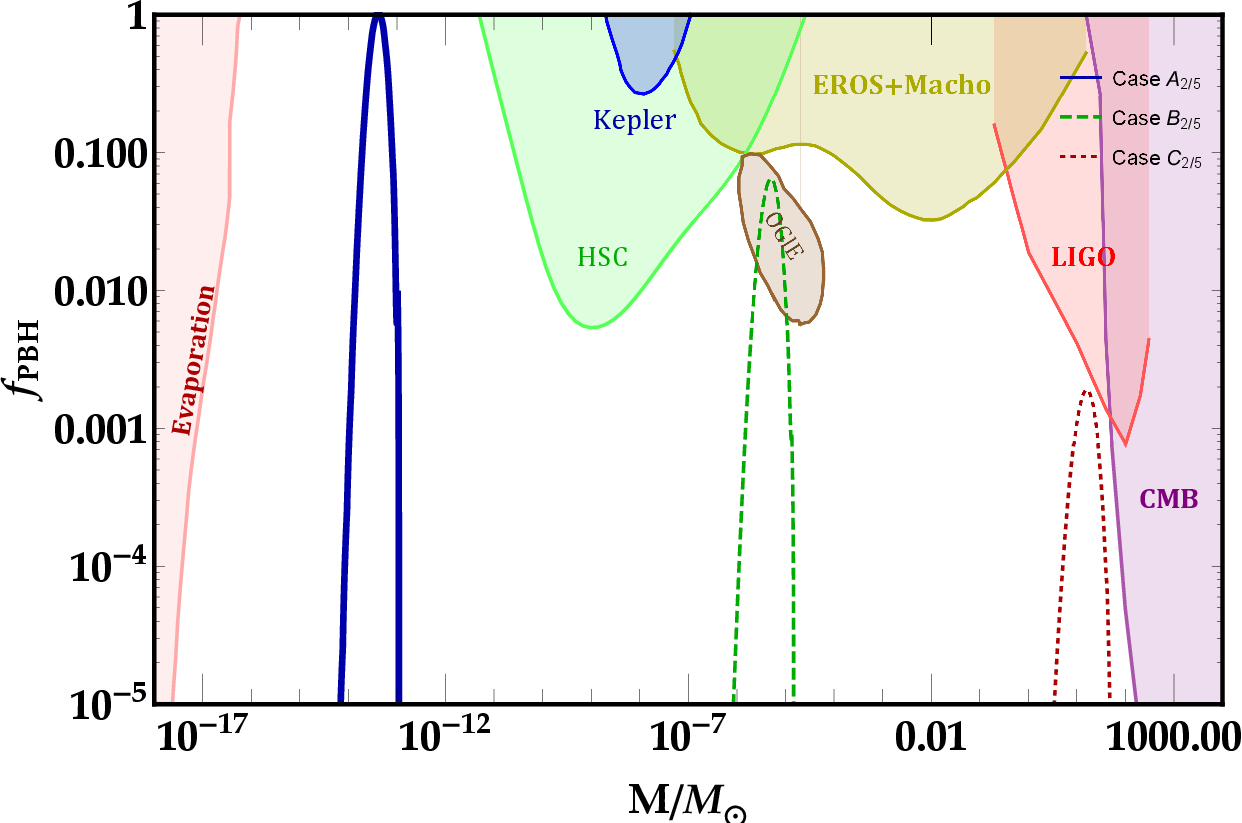}}\\
			\centering
			\subfigure[\label{FPBHs2Per3}]{\includegraphics[width=.49\textwidth]%
				{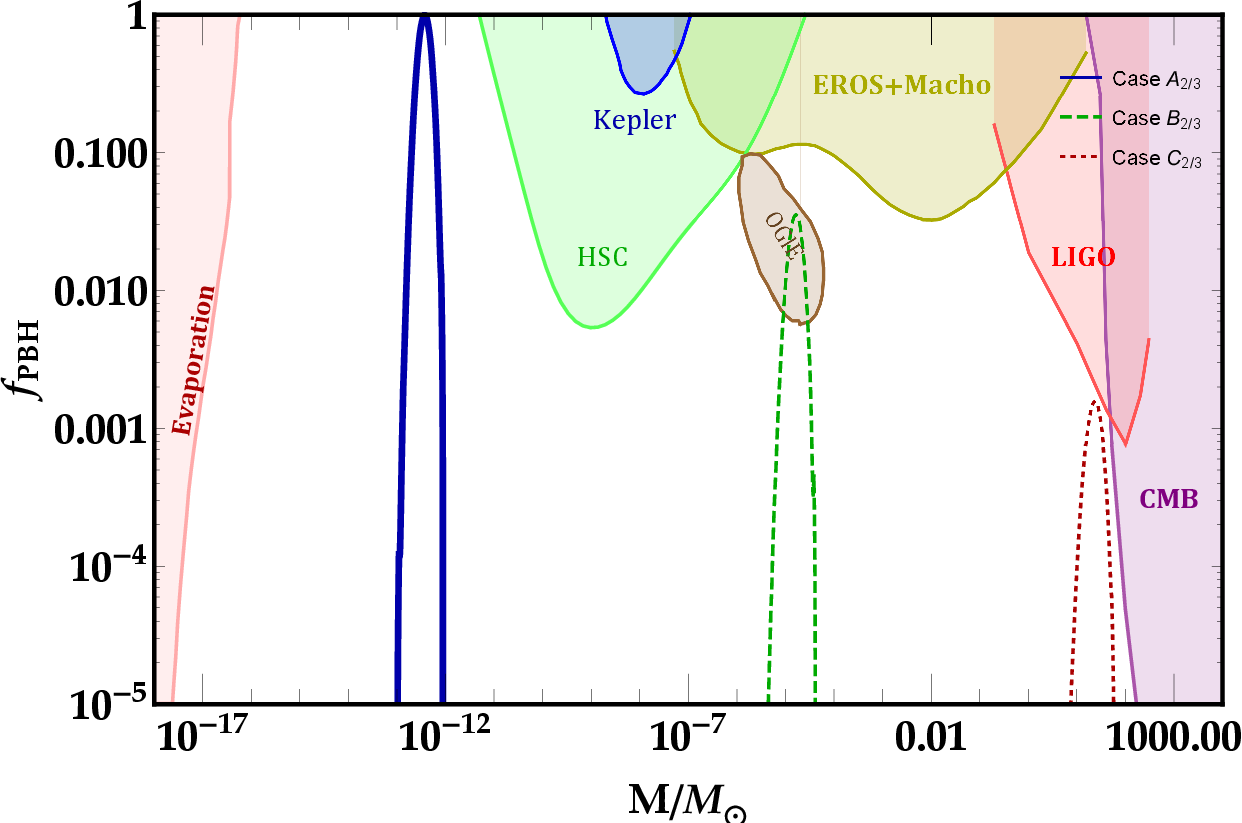}}

		\end{minipage}
		\caption{The abundances of PBHs with respect to their masses $M$ for three parameter sets of potential case with (a) $n=1/3$, (b) $n=2/5$ and (c) $n= 2/3$. The blue, green and red lines in each figure are corresponding to $\rm{Case~}A$, $\rm{Case~}B$ and $\rm{Case~}C$, respectively.
        The shaded regions represent recent observational constraints on the fractional abundance of PBHs. The purple area depicts the restriction on CMB from the signature of spherical accretion of PBHs inside halos \cite{CMB}. The upper bound on the abundance of PBHs is shown by the border of the red shaded region and is derived from the LIGO-VIRGO event consolidation rate \cite{Abbott:2019,Chen:2022,Boehm:2021,Kavanagh:2018}. The brown shaded region represents the authorized region for PBHs abundance due to ultrashort-timescale microlensing events in the OGLE data \cite{OGLE}. The green shaded area corresponds to constraints on microlensing events from collaborations between MACHO \cite{Alcock:2001}, EROS \cite{EORS}, Kepler \cite{Kepler}), Icarus \cite{Icarus}, and Subaru-HSC \cite{subaro}. The pink shaded region delineates constraints related to PBHs evaporation such as extragalactic $\gamma$-ray background \cite{EGG}, galactic center $511$ $keV$ $\gamma$-ray line (INTEGRAL) \citep{Laha:2019}, and effects on CMB spectrum \cite{Clark}.}
\label{FPBHs}
	\end{figure}

Here, our aim is estimating the abundance of PBHs.
It is known that the overdense regions can be created in the RD era due to reentry of the enhanced curvature perturbations to the horizon.
Upon horizon reentry, the resulting overdense regions may gravitationally collapse and generate PBHs.
The mass of PBHs is dependent on the mass of the horizon and is given by
	\ba
	\label{Mpbheq}
	M_{\rm PBH}(k)=\gamma\frac{4\pi}{H}\Big|_{c_{s}k=aH} \simeq M_{\odot} \left(\frac{\gamma}{0.2} \right) \left(\frac{10.75}{g_{*}} \right)^{1/6} \left(\frac{k}{1.9\times 10^{6}\rm Mpc^{-1}} \right)^{-2},
	\ea
where $\gamma= (\frac{1}{\sqrt{3}})^{3}$ \cite{carr:1975} is the collapse efficiency and $g_{*}=106.75$ represents the effective number of relativistic species upon thermalization in the RD era.
The formation rate for PBHs with mass $M(k)$ is calculated using the Press-Schechter theory and assuming Gaussian statistics for the distribution of curvature perturbations \cite{Tada:2019,young:2014} as follows
	\be
	\label{betta}
	\beta(M)=\int_{\delta_{c}}\frac{{\rm d}\delta}{\sqrt{2\pi\sigma^{2}(M)}}e^{-\frac{\delta^{2}}{2\sigma^{2}(M)}}=\frac{1}{2}~ {\rm erfc}\left(\frac{\delta_{c}}{\sqrt{2\sigma^{2}(M)}}\right),
	\ee
where "erfc" stands for the complementary error function, and $\delta_{c}$ represents the threshold value of the density perturbations for PBHs production. Here, we consider $\delta_{c}=0.4$ according to \cite{Musco:2013,Harada:2013}.
In addition  the coarse-grained density contrast $\sigma^{2}(M)$ with the smoothing scale $k$ is defined as follows
	\be
	\label{Sigma}
	\sigma_{k}^{2}=\left(\frac{4}{9} \right)^{2} \int \frac{{\rm d}q}{q} W^{2}(q/k)(q/k)^{4} {\cal P}_{s}(q),
	\ee
%
where ${\cal P}_{s}$ denotes the curvature power spectrum, and $W$ is the Gaussian window function, which is given by $W(x)=\exp{\left(-x^{2}/2 \right)}$.
The ratio of the density parameters of PBHs $(\Omega_{\rm {PBH}})$ to the dark matter $(\Omega_{\rm{DM}})$ at the present time is given by
	\be
	\label{fPBH}
	f_{\rm{PBH}}(M)\simeq \frac{\Omega_{\rm {PBH}}}{\Omega_{\rm{DM}}}= \frac{\beta(M)}{1.84\times10^{-8}}\left(\frac{\gamma}{0.2}\right)^{3/2}\left(\frac{g_*}{10.75}\right)^{-1/4}
	\left(\frac{0.12}{\Omega_{\rm{DM}}h^2}\right)
	\left(\frac{M}{M_{\odot}}\right)^{-1/2}.
	\ee
The current density parameter of dark matter is determined as $\Omega_{\text{DM},0}h^2=0.12$  by Planck 2018 data \cite{akrami:2020}.
Employing ${\cal P}_s$ from solving the MS equation (\ref{M.S}) and using Eqs. (\ref{Mpbheq})-(\ref{fPBH}), we can compute the abundance of the PBHs for all cases listed in Tables~\ref{tab131}, \ref{tab251}, and \ref{tab231}. The numerical results are presented in Tables \ref{tab132}, \ref{tab252}, and \ref{tab232} as well as Fig. \ref{FPBHs}.
	
As depicted in Fig. \ref{FPBHs}, the cases $A_{1/3}$, $A_{2/5}$, and $A_{2/3}$ correspond to PBHs within mass range of $(10^{-16} -10^{-11})M_{\odot}$, can be taken into account for approximately $99\%$ of total dark matter of the universe. The corresponding details are provided in Tables \ref{tab132}, \ref{tab252}, and \ref{tab232}.
For the cases $B_{1/3}$, $B_{2/5}$, and $B_{2/3}$ the abundances of PBHs are $f_{\text{PBH}}^{\text{peak}} = 0.0539$, $f_{\text{PBH}}^{\text{peak}} =0.0644 $, and $f_{\text{PBH}}^{\text{peak}} =0.0346 $, $f_{\text{PBH}}^{\text{peak}}$ respectively. Furthermore, the abundances of these PBHs fall in the allowed region of the observed microlensing events  in the OGLE data.
In addition, the PBHs of cases $C_{1/3}$, $C_{2/5}$, and $C_{2/3}$ are produced with masses around $15.80M_{\odot}$, $15.84M_{\odot}$, and $23.5M_{\odot}$, respectively, with $f_{\text{PBH}}^{\text{peak}} = 0.0017$, $f_{\text{PBH}}^{\text{peak}} =0.0019$, and $f_{\text{PBH}}^{\text{peak}} =0.0015 $. As shown in Fig. \ref{FPBHs}, this class of PBHs is constrained by the upper bound of LIGO-Virgo observations.

\section{Secondary gravitational waves}
\label{sec5}

\begin{figure}[H]
	\begin{minipage}[b]{1\textwidth}
		\centering
		\subfigure[\label{OMEGA1Per3}]{\includegraphics[width=.49\textwidth]%
				{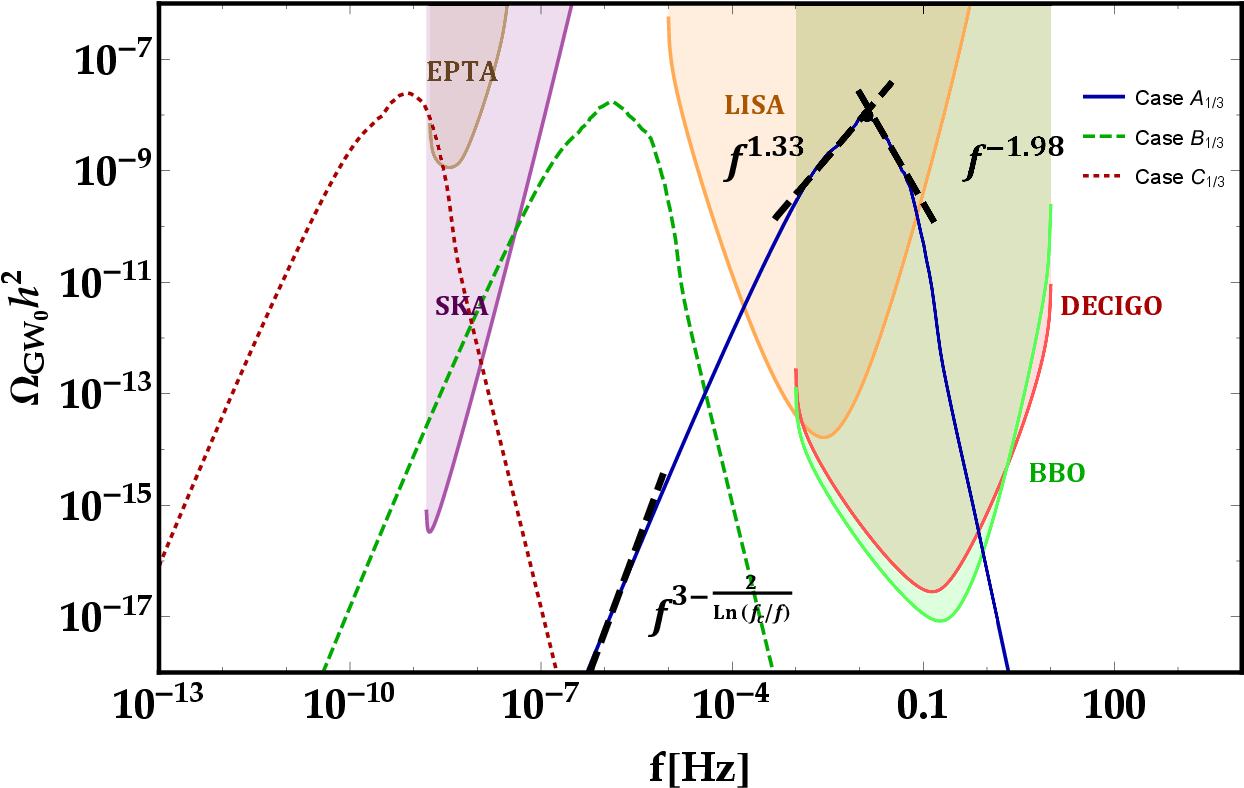}}
		\hspace{.1cm}
		\centering
		\subfigure[\label{OMEGA2Per5}]{\includegraphics[width=.49\textwidth]%
				{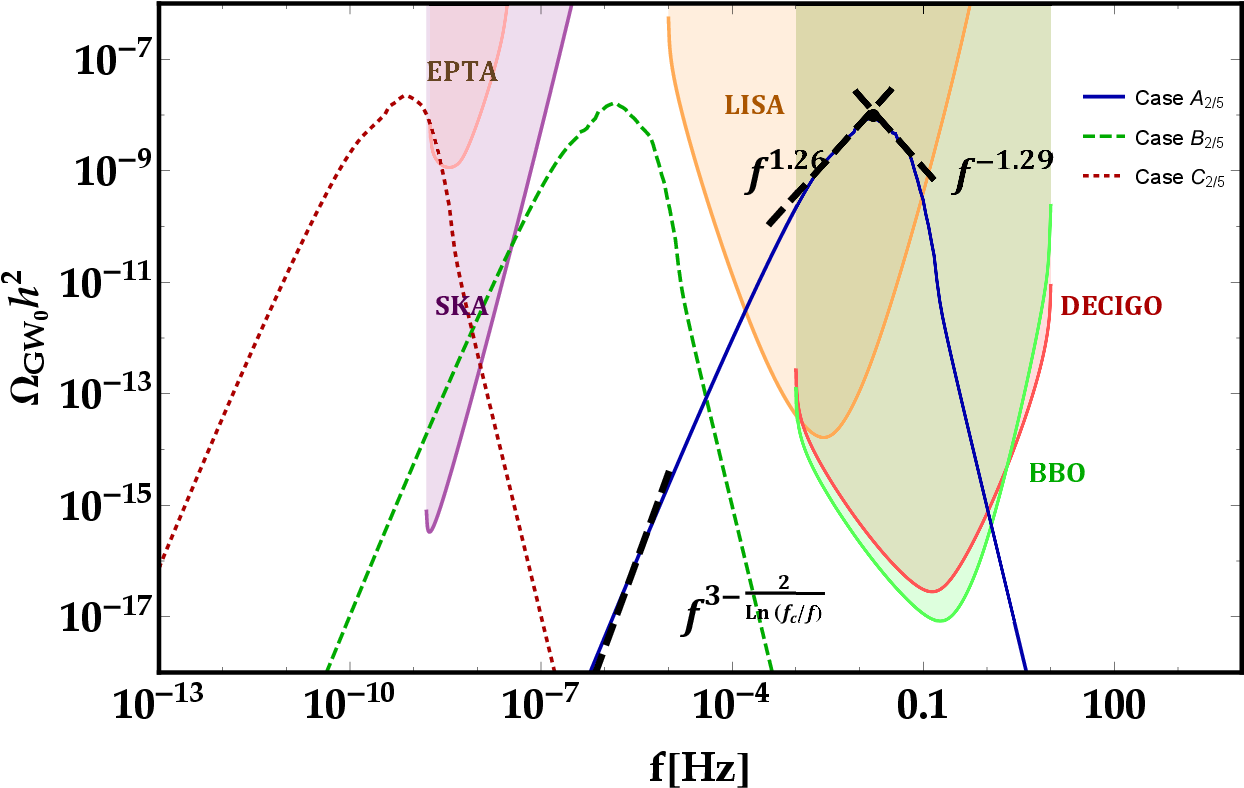}}\\
		\centering
		\subfigure[\label{OMEGA2Per3}]{\includegraphics[width=.49\textwidth]%
				{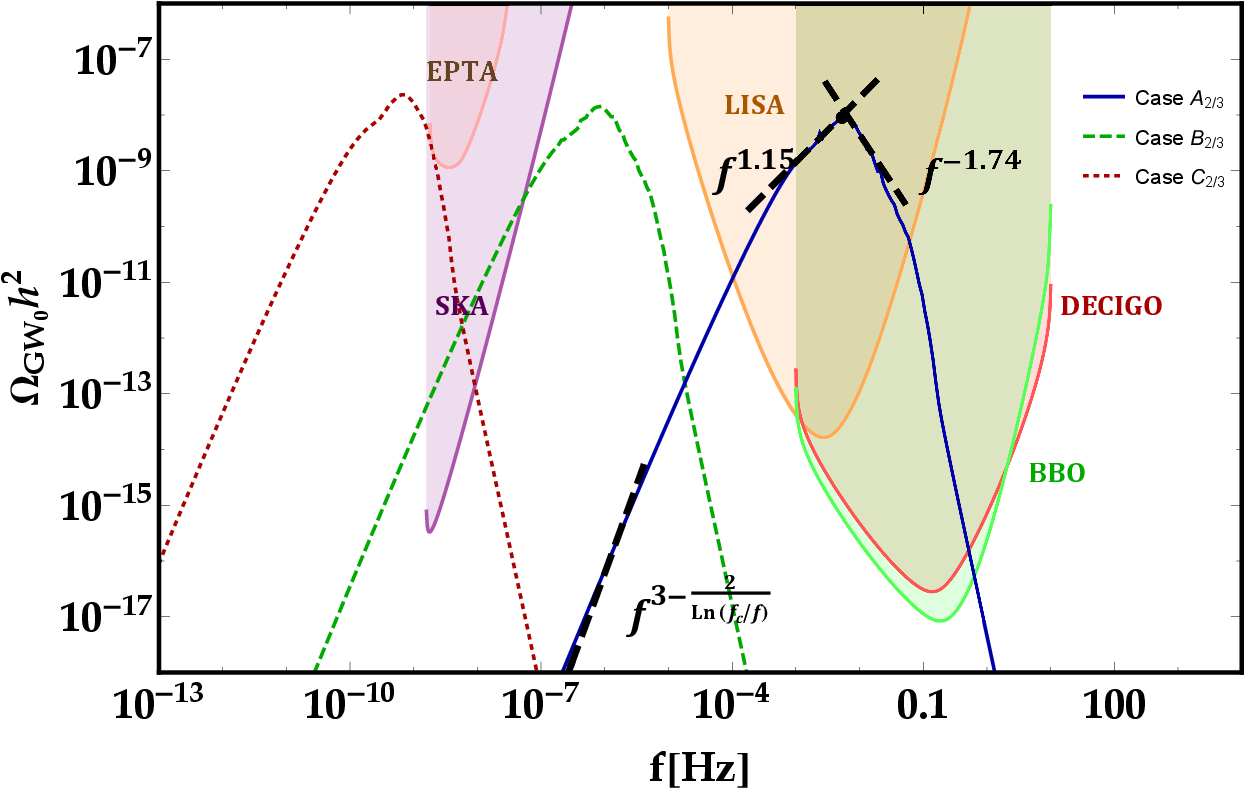}}
			
		\end{minipage}
		\caption{The spectra of present energy density parameter of the secondary gravitational waves $\Omega_{\rm GW_0}$ with regard to frequency pertinent to the fractional power-law potentials with (a) $n=1/3$ (b) $n=2/5$ (c) $n=2/3$ for parameter sets of Table \ref{tab131}, \ref{tab251}, and \ref{tab231}. The blue, green and red lines in each figure are corresponding to $\rm{Case~}A$, $\rm{Case~}B$ and $\rm{Case~}C$, respectively. The accuracy of our predictions can be verified by observatories such as the European PTA (EPTA) \cite{EPTA-a,EPTA-b,EPTA-c,EPTA-d}, the Square Kilometer Array (SKA) \cite{ska,skaCarilli:2004,skaWeltman:2020}, Laser Interferometer Space Antenna (LISA) \cite{ligo-a,ligo-b,lisa,lisa-a}, BBO observatories \cite{Yagi:2011BBODECIGO,Yagi:2017BBODECIGO,Harry:2006BBO,Crowder:2005BBO,Corbin:2006BBO}, and DECIGO \cite{Yagi:2011BBODECIGO,Yagi:2017BBODECIGO,Seto:2001DECIGO,Kawamura:2006DECIGO,Kawamura:2011DECIGO}.
		}\label{OMEGA}
	\end{figure}

Secondary gravitational waves production is a consequence of re-entering the enhanced amplitude of primordial curvature perturbations  to the horizon during the RD era. The emitted GWs can be detected by the various detectors with different sensitivity ranges.
Here, we discuss the production of secondary GWs in GB framework with a fractional power-law potentials.
At the end of inflation, the inflaton field decays and transforms into light particles, which heat the universe and initiate the RD epoch.
Hence, the  impact of the inflaton field on the cosmic evolution is negligible during RD epoch. Consequently, the standard Einstein formulation can be used to investigate the production of secondary GWs during this era.
The energy density parameter of secondary GWs has been studied in \cite{Ananda:2007,Baumann:2007,Kohri:2018} and it is calculated as follows
\ba
\label{OGW}
&\Omega_{\rm{GW}}(\eta_c,k) = \frac{1}{12} {\displaystyle \int^\infty_0 dv \int^{|1+v|}_{|1-v|}du } \left( \frac{4v^2-(1+v^2-u^2)^2}{4uv}\right)^2\mathcal{P}_{s}(ku)\mathcal{P}_{s}(kv)\left( \frac{3}{4u^3v^3}\right)^2 (u^2+v^2-3)^2\nonumber\\
&\times \left\{\left[-4uv+(u^2+v^2-3) \ln\left| \frac{3-(u+v)^2}{3-(u-v)^2}\right| \right]^2  + \pi^2(u^2+v^2-3)^2\Theta(v+u-\sqrt{3})\right\}\;,
\ea
where, $\Theta$ denotes the Heaviside theta function, and $\eta_{c}$ represents the time at which the growth of $\Omega_{\rm{GW}}$ ceases. According to the following equation, the present value of the energy spectra is related to the energy spectra at $\eta_{c}$ \cite{Inomata:2019-a}
\be
\label{OGW0}
\Omega_{\rm GW_0}h^2 = 0.83\left( \frac{g_{*}}{10.75} \right)^{-1/3}\Omega_{\rm r_0}h^2\Omega_{\rm{GW}}(\eta_c,k)\,,
\ee
where the present value of radiation density parameter is denoted by $\Omega_{\rm r_0}h^2\simeq 4.2\times 10^{-5}$, and $g_{*}\simeq106.75$ is the effective degrees of freedom in the energy density at $\eta_c$.
It is worth noting that, the frequency is related to the wavenumber as follows
\be
\label{k_to_f}
f=1.546 \times 10^{-15}\left( \frac{k}{{\rm Mpc}^{-1}}\right){\rm Hz}.
\ee
In our study, we utilize the accurate scalar power spectrum estimated in previous sections and Eqs.~(\ref{OGW})-(\ref{k_to_f}) to obtain the present energy spectra of GWs associated with PBHs for all cases listed in Tables \ref{tab131}, \ref{tab251}, and \ref{tab231}.
Our predicted results, along with the sensitivity curves of various GWs observatories, are shown in Fig. \ref{OMEGA}.
For cases $A$ (including $A_{1/3},~A_{2/5},$ and $A_{2/3}$), $B$ (including $B_{1/3},~B_{2/5},$ and $B_{2/3}$), $C$ (including $C_{1/3},~C_{2/5},$ and $C_{2/3}$), the spectra of $\Omega_{\rm GW_0}$ has located at different frequencies with similar magnitudes of order $10^{-8}$.
Regarding case $A$ associated with asteroid-mass PBHs, the peaks of $\Omega_{\rm GW_0}$ spectra are present in the mHz and cHz  frequency range, which can be detected by observatories such as  LISA\cite{ligo-a,ligo-b,lisa,lisa-a}, BBO  \cite{Yagi:2011BBODECIGO,Yagi:2017BBODECIGO,Harry:2006BBO,Crowder:2005BBO,Corbin:2006BBO}, and DECIGO \cite{Yagi:2017BBODECIGO,Seto:2001DECIGO,Kawamura:2006DECIGO,Kawamura:2011DECIGO}. For cases $B$ and $C$ corresponding to earth-mass and stellar-mass PBHs, respectively, the peaks of $\Omega_{\rm GW_0}$ are localized at frequencies around  $f_{c}\sim10^{-6}\text{Hz}$ and $f_{c}\sim10^{-9}\text{Hz}$, which fall in the sensitivity range of the SKA \cite{ska,skaCarilli:2004,skaWeltman:2020} observatory.

The predicted $\Omega_{\rm GW_0}$ spectra for all cases of our model intersect the sensitivity curves of various GWs observatories. Hence, the validity of this model can be evaluated using data of these observatories in the future.
Finally, we examine the behaviour of the $\Omega_{\rm GW_0}$ spectra at different frequency ranges near the peak position. It has been confirmed that near the peak position, the current density parameter spectra of secondary GWs exhibit a power-law behaviour with respect to frequency, i.e. $\Omega_{\rm GW_0} (f) \sim f^{n} $ \cite{Fu:2019vqc,Fu:2020lob,Bagui:2021dqi,Xu,Kuroyanagi}. As displayed in Table \ref{tableGWs}, our model exhibits the expected behaviour near the peak of  $\Omega_{\rm GW_0}$.
We provide the exact values of the critical frequencies and peak heights for all cases in Table \ref{tableGWs}.
Furthermore, our obtained results in the infrared regime $f \ll f_c$ confirm the analytical relation $\Omega_{\rm GW_0}\sim f^{3-2/\ln\left(f_{c}/f\right)}$ derived by \cite{Yuan:2020,Sasaki:2020} as shown in Fig. \ref{OMEGA}.
	
    \begin{table}[H]
        \centering
        \caption{The frequencies and heights of the peaks of the  $\Omega_{\rm GW_0}$ spectra  beside power index $n$ in frequency ranges  $f\ll f_{c}$, $f<f_{c}$ and  $f>f_{c}$ for Cases $A_{\rm 1/3}$, $A_{\rm 2/5}$, and $A_{\rm 2/3}$.}
		\scalebox{1}[1] {
			\begin{tabular}{cccccc}
				\hline
				\#  & $\qquad\qquad$ $f_{c}$ $\qquad\qquad$ & $\quad$ $\Omega_{\rm GW_0}\left(f_{c}\right)$ $\quad$ & $\quad$ $n_{f\ll f_{c}}$ $\quad$ & $\quad$ $n_{f<f_{c}}$ $\quad$ & $\quad$ $n_{f>f_{c}}$\tabularnewline
				\hline
				\hline
				$\rm{Case~}A_{\rm 1/3}$ & $1.265\times10^{-2}$ & $9.867\times10^{-9}$ & $3.0$ & $1.33$ & $-1.98$ \tabularnewline
				\hline
				$\rm{Case~}A_{\rm 2/5}$ & $1.633\times10^{-2}$ & $9.705\times10^{-9}$ & $3.0$ & $1.26$ & $-1.29$ \tabularnewline
				\hline
				$\rm{Case~}A_{\rm 2/3}$ & $5.254\times10^{-3}$ & $9.110\times10^{-9}$ & $2.98$ & $1.15$ & $-1.74$ \tabularnewline
				\hline
			\end{tabular}
		}
		\label{tableGWs}
	\end{table}

\section{Conclusions}
\label{sec6}
In this paper, we have investigated the possibility of PBHs production in an inflationary model where the scalar field is coupled to the GB term.
Additionally, we have considered the fractional power-law potentials in this study.
In the presence of the GB coupling term, inflaton passes through the vicinity of a non-trivial fixed point and experiences an USR epoch. As a result, the primordial curvature perturbations could enhance during this era.

Here, we have considered three type of the fractional power-law potentials in the form of $($$\phi^{1/3}$, $\phi^{2/5}$ and $\phi^{2/3}$$)$, which are ruled out in the standard inflationary model by Planck 2018 \cite{akrami:2020}.
This attribute prompted us to revise them in GB framework to achieve a viable inflationary model.
By using the GB coupling term and fine-tuning the model parameters, namely $V_0$, $\phi_{c}$, $\xi_{1}$, $\xi_{0}$ (see Tables \ref{tab131}, \ref{tab251}, and \ref{tab231}), we have successfully reconciled the fractional power-law potentials with the results obtained from the CMB measurements.
Furthermore, an epoch of USR inflation can be achieved at small scales. During this period, the inflaton experiences a significant deceleration, which is necessary for the PBHs formation.

Using the exact solution of background equations, we demonstrated the behavior of inflaton field $\phi$, Hubble parameter $H$, slow-roll parameters ($\e_1$ and $\e_2$), $c_s^{2}$ and $c_t^{2}$ in Figs. \ref{Fig131}, \ref{Fig251} and \ref{Fig231} with respect to $e$-fold number.
One can see from Figs. \ref{Fig131}, \ref{Fig251} and \ref{Fig231} that during the USR phase, the slow-roll conditions are confirmed by the first parameter ($\e_1\ll1$), but broken by the second one ($\left|\e_2\right| \geq 1$).
In this regards, we obtained the numerical solution of the MS equation (\ref{M.S}) to calculate the exact values of the scalar power spectra ${\cal P}_s$ for all cases of Tables \ref{tab132}, \ref{tab252}, and \ref{tab232}.
Additionally, Figs. \ref{Fig132}, \ref{Fig252} and \ref{Fig232} show that the obtained power spectra are compatible with Planck 2018 data on large scales and have  peaks of  sufficient heights to generate PBHs on small scales.
According to the predictions of our model, the values of $n_s$ and $r$ for all cases of $\phi^{1/3}$ and $\phi^{2/5}$  are consistent with the $68\,\%\,\rm CL$ of the Planck  2018  TT, TE, EE + lowE + lensing + BK15 +BAO data, and the values of $n_s$ and $r$ for all Cases of $\phi^{2/3}$ gratify the $95\,\%\,\rm CL$ \cite{akrami:2020}.
Consequently, by using the GB framework, our investigation revived three fractional power-law potentials $($$\phi^{1/3}$, $\phi^{2/5}$ and $\phi^{2/3}$$)$, that are previously ruled out in the standard inflationary model.

In the following by using the scalar power spectrum, we achieved detectable PBHs abundance with masses  of ${\cal O}(10)M_\odot$ for cases  $C_{\rm 1/3}$, $C_{\rm 2/5}$, and $C_{\rm 2/3}$, which fall in the stellar-mass category and are appropriate for describing the LIGO-Virgo events.
In addition, we have identified PBHs with mass scales of ${\cal O}(10^{-6} - 10^{-5})M_\odot$ for cases $B_{\rm 1/3}$, $B_{\rm 2/5}$, and $B_{\rm 2/3}$.
The abundance of this class of PBHs is of ${\cal O}(10^{-2})$ and could explain the microlensing events in OGLE data.
The most interesting class of our results is related to the case $A$ including $A_{\rm 1/3}$, $A_{\rm 2/5}$, and $A_{\rm 2/3}$, corresponding to PBHs in the asteroid-mass range.
The  first two cases yield PBHs with mass scales of ${\cal O}(10^{-14})M_\odot$, while the latter case produces  PBHs with  mass of ${\cal O}(10^{-13})M_\odot$. In this class, the predicted PBHs could contain $99 \%$ of the total dark matter of the universe, as presented in Tables \ref{tab132}, \ref{tab252}, and \ref{tab232}.

Finally, we investigated the generation of induced GWs following the PBHs formation for all cases of our model.
The estimated current density parameter spectra ($\Omega_{\rm GW_0}$) indicate that while each case exhibit a peak at a different frequency, all cases have approximately identical heights of ${\cal O}(10^{-8})$ (see Table \ref{tableGWs}).
The peaks of $\Omega_{\rm GW_0}$ for Cases $C_{\rm 1/3}$, $C_{\rm 2/5}$, and $C_{\rm 2/3}$ are located at frequencies around ${\cal O}(10^{-10})\text{Hz}$ and can be tested by the SKA detector.
Besides, the spectra of $\Omega_{\rm GW_0}$ for Cases $B_{\rm 1/3}$, $B_{\rm 2/5}$, and $B_{\rm 2/3}$ have peaks localized in the ${\cal O}(10^{-6})\text{Hz}$ bands, which could be observed by SKA observatory.
Moreover, the spectra of $\Omega_{\rm GW_0}$ for Cases $A_{\rm 1/3}$, $A_{\rm 2/5}$, and $A_{\rm 2/3}$ exhibit peaks about ${\cal O}(10^{-2})\text{Hz}$ bands, which are detectable by LISA, BBO, and DECIGO observatories (see Fig.~\ref{OMEGA}). Consequently, the validity of our model can be assessed by examining GWs using the extracted data of these detectors. In addition, we have denoted that the $\Omega_{\rm GW_0}$ spectra behaves as a power-law function of frequency ($\Omega_{\rm GW_0)}(f)\sim f^{n}$). We have calculated that $\Omega_{\rm GW_0} \sim f^{1.33}$, $\Omega_{\rm GW_0} \sim f^{1.26}$, and $\Omega_{\rm GW_0} \sim f^{1.15}$ for Cases $A_{\rm 1/3}$, $A_{\rm 2/5}$, and $A_{\rm 2/3}$, respectively, in the $f<f_{c}$ region.
Similarly, we calculated $\Omega_{\rm GW_0} \sim f^{-1.98}$, $\Omega_{\rm GW_0} \sim f^{-1.29}$, and $\Omega_{\rm GW_0} \sim f^{-1.74}$ for Cases $A_{\rm 1/3}$, $A_{\rm 2/5}$, and $A_{\rm 2/3}$, respectively, in the frequency range of $f>f_{c}$.
In the infrared domain  $f\ll f_{c}$, the  power index has been specified as $n=3-2/\ln(f_c/f)$ (see Table \ref{tableGWs}), which is consistent with the analytical obtained results of \cite{Yuan:2020,Sasaki:2020}.


	
\end{document}